\documentclass[lettersize,journal]{IEEEtran}
\usepackage{amsmath,amsfonts}
\usepackage{algorithmic}
\usepackage{algorithm}
\usepackage{array}
\usepackage[caption=false,font=normalsize,labelfont=sf,textfont=sf]{subfig}
\usepackage{textcomp}
\usepackage{stfloats}
\usepackage{bibentry}
\usepackage{url}
\usepackage{verbatim}
\usepackage{graphicx}
\usepackage{cite}
\usepackage{multirow}
\usepackage{balance}
\usepackage{subfloat}
\usepackage{threeparttable}
\usepackage{makecell}
\usepackage{stfloats}

\def\BibTeX{{\rm B\kern-.05em{\sc i\kern-.025em b}\kern-.08em
		T\kern-.1667em\lower.7ex\hbox{E}\kern-.125emX}}
\usepackage{balance}
\begin{document}
		\title{Vision-Aided     Channel   Prediction Based on Image Segmentation at Street Intersection Scenarios}
	\author{Xuejian Zhang,~\IEEEmembership{Graduate Student Member,~IEEE,} Ruisi He,~\IEEEmembership{Senior Member,~IEEE,}   Mi Yang,~\IEEEmembership{Member,~IEEE,} 
		Ziyi Qi, Zhengyu Zhang,~\IEEEmembership{Graduate Student Member,~IEEE,}  
		Bo Ai,~\IEEEmembership{Fellow,~IEEE,} Zhangdui Zhong,~\IEEEmembership{Fellow,~IEEE} 
	\thanks{
Part of this paper was submitted to the 
 14th International Symposium on Antennas, Propagation and EM Theory (ISAPE 2024) \cite{ISAPE}.

X. Zhang, R. He, M. Yang, Z. Qi,  Z. Zhang,  B. Ai, and Z. Zhong are with the School of Electronics and Information Engineering and the Frontiers Science Center for Smart High-speed Railway System, Beijing Jiaotong University, Beijing 100044, China
(email: 23115029@bjtu.edu.cn; ruisi.he@bjtu.edu.cn;  myang@bjtu.edu.cn; 22115006@bjtu.edu.cn; 21111040@bjtu.edu.cn;  boai@bjtu.edu.cn; zhdzhong@bjtu.edu.cn).
%


}
}
	\maketitle

\begin{abstract}
Intelligent vehicular communication with vehicle-road collaboration capability is a key technology enabled by 6G, and the integration of various visual sensors  on vehicles and infrastructures plays a crucial role. 
Moreover, accurate  channel prediction is foundational to realizing intelligent vehicular communication. 
Traditional methods are still limited by the inability to balance accuracy and operability based on  substantial spectrum resource consumption and highly refined description of environment. 
Therefore, leveraging  out-of-band information introduced by visual sensors   provides a new solution and  is increasingly applied across various communication tasks.
In this paper,
we   propose a  computer vision (CV)-based  prediction model  for    vehicular communications, realizing accurate channel characterization prediction including path loss, Rice K-factor and delay spread    based on image segmentation.
First, we conduct  extensive vehicle-to-infrastructure measurement campaigns, collecting channel and visual data from various street intersection scenarios. 
The image-channel dataset is generated after a series of data post-processing steps. 
Image data consists of individual segmentation of  target user  using  YOLOv8 network. 
Subsequently,  established dataset is used to train and test  prediction network ResNet-32,  where   segmented images serve as   input of network, and various channel characteristics are treated as   labels or target outputs of  network. 
Finally, self-validation and cross-validation experiments are   performed.
The results indicate that models trained with  segmented images achieve high prediction accuracy and remarkable generalization performance across different streets and target users.
The model proposed in this paper offers novel solutions for achieving intelligent channel prediction in vehicular communications.

\end{abstract}


\section{Introduction}
\IEEEPARstart{I}{n} 
recent years, 5G and 6G communication technologies have rapidly developed worldwide, with the design and construction of next-generation communication systems receiving extensive attention \cite{he2024}. 
At the same time, the development of artificial intelligence (AI) theory and technological innovations has significantly accelerated the transition of wireless communications from digitization and networking to intelligence \cite{Huang2022}. 
ITU-R has released a recommendation outlining the overall goals for IMT-2030 (6G), which includes the integration of AI and communications as one of  six major scenarios for future 6G \cite{itu1,itu2}. 
COST CA20120 Action \cite{COST, he2024cost} also focuses on advancing intelligent wireless communication technologies in 6G to support seamless and inclusive interactions. 
It can be seen that deep convergence between AI and communication technologies has become an inevitable trend in future 6G \cite{Yang2021}. 

As one of the pivotal applications of 6G, intelligent vehicular communications is progressively steering towards achieving ultra-low latency and enhanced vehicle-road collaboration,  such as autonomous driving and intelligent transportation \cite{hrs2020}. 
The joint utilization of multimodal sensors integrated into both vehicles and roadside infrastructures, such as radar \cite{deepsense}, LiDAR \cite{QI2023241}, and depth/RGB cameras \cite{Nishio2019,Charan2024}, plays a crucial role in ensuring the reliability and effectiveness of vehicular communication system. 
Furthermore, accurate channel prediction is also foundational to  realizing intelligent vehicular communications  \cite{Zhang2024}. 
Currently, mainstream   methods primarily employ conventional statistical and electromagnetic techniques to predict and model channel characteristics \cite{yangmi}. 
They rely on  occupation of spectrum resources or  detailed characterization of  propagation environment.
However, these methods face significant challenges in balancing high accuracy, ease of implementation, and strong generalization.
As the complexity of 6G communication scenarios and channel characteristics continues to increase, traditional statistical and deterministic channel models are becoming increasingly intricate, while the adaptability to dynamic scenarios and the self-evolution of models remain major bottlenecks  \cite{huangchen}.
The rapid advancement of computer vision (CV) technologies, coupled with the integration of visual sending data through  extensive deployment of multimodal sensors in vehicular communications, offers a promising solution to this challenge \cite{APS2024}.

Channel  prediction in vehicular communications is one of the potential typical  applications of CV-aided technologies \cite{he2019applications}. 
On the one hand, 
visual sensing data  is out-of-band information and do not occupy frequency resources, which can reduce communication overhead and improving spectral efficiency \cite{tian}. 
Furthermore, with the gradual development of  a wide range of sensors, making it much easier to obtain visual data. 
On the other hand, 
there is an inherent correlation between wireless channel   and visual data \cite{Nishio2021}. 
Wireless channel is determined by propagation environment, which can be recorded by visual sensors \cite{zzy2023}. 
Visual data can reflect information such as the sizes, shapes, and locations of scatterers and potential target users in  propagation environment \cite{CJE}. 
By leveraging  CV technology and deep learning networks,  visual data acquired through vehicular communications, such as vehicle-to-vehicle (V2V) and vehicle-to-infrastructure (V2I) interactions, can be effectively utilized to characterize and predict wireless propagation dynamics in real-time. 
This approach not only avoids consuming additional communication resources but also ensures high accuracy and robust generalization.
  
Many studies have explored channel prediction using visual   data. 
\cite{wcl} design a novel convolutional neural network (CNN) to extract environmental features from satellite images for predicting PL  over large areas. 
The influence of environmental granularity in satellite images on  accuracy of PL prediction is discussed in \cite{qzc}. 
Based on geometric feature maps, a prediction model for received power in indoor environments is proposed in \cite{Seretis2022}. 
\cite{lee2019} utilizes three-dimensional (3D) building maps as input to CNN for PL prediction in urban areas. 
3D building maps and point clouds are combined  to achieve PL prediction in urban canyon in \cite{Gupta2022}. 
Based on depth images, received power prediction  for millimeter-wave (mmWave) links is achieved in \cite{Nishio2019}. 
These studies demonstrate the widespread application of diverse visual data combined with AI technologies in channel prediction. 
However, the above studies focus primarily on predicting PL and received power and do not extend to other typical channel characteristics. 
What's more, they predominantly involve static or delayed predictions, which limits their applicability to dynamic environments such as vehicular communications and restricts their ability to provide real-time predictions.

There are also some studies on CV-aided wireless communications for vehicular communications. 
\cite{alrabeiah2020viwi,Alrabeiah2020,Charan2021} obtain   simulated time series RGB images from   base station and use  deep neural networks to implement beam prediction, blockage prediction and proactive handoff in mmWave multiple-input multiple-output (MIMO) networks.
\cite{Xu2023_2,Xu2023,Feng2024} use vision-aided methods with RGB images to achieve beam alignment, multi-user matching and resource allocation in mmWave vehicle communications and smart factory scenarios,  respectively.
To achieve beam and blockage prediction better  in mmWave MIMO networks,
\cite{env, feifei2023, feifeigao2023} put  forward  concepts of environment semantic from different perspectives, which is defined  the key environmental information extracted from images.
Although the aforementioned studies are application cases in dynamic scenarios, their task objectives are mainly beam prediction, blocking prediction, etc., rather than channel characterization prediction. There is a significant distinction between these tasks and channel prediction.

In summary, there is a lack of research that leverages CV  technology and RGB images to assist in predicting typical channel characteristics in vehicular communications. 
On one hand, most studies that use non-real-time visual data, such as satellite images and 3D building maps, only study the prediction of PL and received power, and can hardly support real-time vehicular communications. 
On the other hand, while some studies use dynamic visual sensing data in V2V or V2I scenarios, these efforts are mainly directed towards assisting with beam and blockage prediction, without addressing the prediction of channel characteristics.
  
To address  the issues discussed above, we presented some preliminary experimental results in our previous work of \cite{ISAPE}.
In this paper,  CV-based method and model for channel prediction are proposed in  vehicular communications, typical channel characteristics predicted include path loass (PL), Rice K-factor and root mean square (RMS) delay spread.
Instance segmentation method for RGB images   and two widely used deep neural networks, i.e. YOLOv8 and ResNet-34,  are used.
The proposed model is highly applicable in V2I scenarios. 
It requires only the dynamic moving RGB images of target user, captured at base station, and then input  separately segmented images of  target user into  prediction network
to predict   channel characteristics, thereby realizing the association between environmental sensing information and wireless channel.
The main contributions of this paper are summarized as follows.

\begin{itemize}
	\item[$\bullet$]  
	Channel measurement campaigns in  two street intersection scenarios are widely carried out.
    The methods for acquiring channel and visual data in V2I scenarios, along with the process of one-to-one matching to generate image-channel datasets, are presented in detail.
	\item[$\bullet$]  
	A vision-aided channel    prediction model based on image segmentation methods for  V2I  channel is proposed. 
	The model framework and implementation steps are presented in detail. 
	In this model, accurate channel characteristics including PL, Rice K-factor and  RMS delay spread can be predicted solely by inputting RGB images containing the target user.
	\item[$\bullet$]  The prediction accuracy and generalization performance of the model are verified on the measured datasets, including scenario self-validation and cross-validation. 
    The proposed model demonstrates high prediction accuracy and exhibits robust generalization performance across diverse streets and target users.

\end{itemize}

The remainder of this paper is organized as follows. 
Section II describes the system model and framework of proposed model. 
Section III   introduces   experiment  setup, including measurement campaign, data post-processing and model training. 
Then in Section IV, a series of experiments about performance validation and analysis for   proposed model is presented. 
Finally, Section V draws the conclusions.

\section{Vision-aided Channel Prediction Model}  

\subsection{System Model}
In this study, we consider a typical  V2I  communication scenario, where a base station (BS) is located at an intersection along the roadside, as shown in Fig. \ref{model}. 
Furthermore, BS is equipped with an omnidirectional transmitting antenna and a standard-resolution RGB wide-angle camera, which together enable communication and environmental monitoring. 
The mobile user (i.e.,  vehicle) is equipped with a single omnidirectional antenna capable of receiving  vector channel sounding signals transmitted by BS. 
Without considering  MIMO,  channel impulse response $h(t,\tau )$ can be expressed as
\begin{equation}
	\label{math1}
     h(t,\tau ) = \sum\limits_{l = 1}^L {{\alpha _l}{e^{ - j\phi {}_l}}\delta (t - {\tau _l})}, 
\end{equation}
where $t$ is  index of  time snapshot, $ L$ are the number  of   rays, and $\tau _l $ is the delay of the $ L$-th path. 
What's more, $ \delta ( \cdot )$ is the Dirac delta function, and $ \phi {}_l $ is the phase of paths that is assumed to be described by statistically independent random variables uniformly distributed over $[0,2\pi )$.
Performing Fourier transform on $h(t,\tau )$ can obtain the channel transfer function $H(f)$.
Thus we can obtain  channel characteristics, typical examples of which include PL,  Rice K-factor, PDP, RMS delay spread, etc.

PL  are the large-scale propagation characteristics, 
which reflect the attenuation of radio waves over larger distances (tens to hundreds of times  wavelength) \cite{Zhang2023}.
which can be determined by averaged channel gain as follows
\begin{equation}
	\label{math2}
	PL =  - 10{\log _{10}}(\frac{1}{W}\sum\limits_{T = t - \frac{W}{2}}^{t + \frac{W}{2} - 1} {\sum\limits_{\tau  = 1}^{{N_f} } {{{\left| {h(T,\tau )} \right|}^2}} } ),
\end{equation}
where ${N_f}$ is the number of  frequency points, $PL$ is the path loss in dB scale,
and $W$ is a $40\lambda$ sliding window, $\lambda$ denotes wavelength.

Rice K-factor represents the ratio of the power of line-of-sight (LoS) component and the power of non-line-of-sight (NLoS) component in the channel \cite{MiYang2023}.
The calculation formula is as follow
\begin{equation}
	\label{math3}
	K = \frac{{{{\left| {{h_{LOS}}} \right|}^2}}}{{\sum\limits_\tau  {{{\left| {{h_{NLOS}}} \right|}^2}} }},
\end{equation}
where $h_{LOS}$ is LOS component, and $h_{NLOS}$ is NLOS component. 
All valid multipaths can be identified well  using multipath discrimination algorithm \cite{fyy}. 
The multipath with the highest power in single snapshot is designated as  $h_{LOS}$, while the remaining valid multipaths are classified as $h_{NLOS}$.

Power delay profile (PDP) is extensively employed to characterize the power levels of received paths with propagation delays and to describe the distribution of multi-path components  in measured environments.
The instantaneous PDP is denoted as
\begin{equation}
\label{math4}
P(t,\tau ) = {\left| {h(t,\tau )} \right|^2}.
\end{equation}

RMS DS  is the square root of the second central moment of average PDPs (APDPs), obtained by averaging with a sliding window of $40\lambda$, and is widely used to characterize the delay dispersion of channels. 
It is defined as
\begin{equation}
\label{math5}
{\tau _{rms}}(d) = \sqrt {\frac{{\sum\nolimits_l {APDP(d,{\tau _l})\tau _l^2} }}{{\sum\nolimits_l {APDP(d,{\tau _l})} }} - {{(\frac{{\sum\nolimits_l {APDP(d,{\tau _l}){\tau _l}} }}{{\sum\nolimits_l {APDP(d,{\tau _l})} }})}^2}}, 
\end{equation}
where $\tau _l$ represents the delay of the $l$th path and $APDP(d,{\tau _l})$ describes the corresponding power with $\tau _l$ measured at location $d$.

\subsection{Problem Formulation}
Given the system model in Section II-A, at any given time instant $t$,
the task of channel prediction is to capture the evolving trends of channel characteristics in order to inform key aspects of communication systems, such as network planning, base station placement, power control, and error correction algorithms, etc. of communication systems. 
Traditional approaches typically rely on channel models, statistical methods based on channel measurements, or deterministic methods such as ray tracing and parabolic equations, to predict channel propagation characteristics. 
However,  channel models established based on traditional methods often struggle to balance accuracy with computational complexity, lack generalization and adaptability across diverse scenarios, 
and fail to  leverage the characteristics of  propagation environment.

A promising alternative is the application of artificial intelligence, particularly computer vision, to perceive  propagation environment and thereby assisting in  channel prediction. 
Visual data, which can be easily acquired and does not consume valuable spectrum resources, represents a valuable out-of-band resource. 
In recent years, research on exploring the use of visual data to assist wireless communication has gained traction, particularly in areas such as blockage prediction, beamforming prediction, and  proactive handoff in millimeter-wave (mmWave) and terahertz (THz) scenarios.
In this paper,  we propose the use of additional visual data—specifically, RGB images captured by camera installed on BS—in conjunction with computer vision and deep learning to predict channel propagation characteristics.
Notely, instead of using   raw images, we follow the suggestion of \cite{env} and use RGB images after instance segmentation.

Formally, we define $ {\rm X}[t] \in \mathbb{R} {^{W \times H \times C}} $ as the corresponding RGB image  at time $t$, where $W$, $H$, $C$ are the width, height, and the number of color channels of the image.
Let, $ {\rm Z}[t]$ represent RGB image after instance segmentation. 
The objective of channel characteristic prediction task is to find a mapping function ${f_\Theta }$ that utilizes $ {\rm Z}[t]$ to predict channel characteristics $\widehat c[t] \in \Omega $,  $\Omega  = [PL,K,{\tau _{rms}},...]$. 
The mapping function can be formally expressed as
\begin{equation}
	\label{math6}
	{f_\Theta }:{\rm Z}[t] \to \widehat c[t].
\end{equation}

In this paper, we develop a deep learning model to learn this prediction function ${h_\Theta }$.
Let $D = \left\{ {\left( {{Z_m},c_m^ \star } \right)} \right\}_m^M$ represent available dataset consisting of ``segmented RGB image-channel characteristic'' pairs is collected from wireless propagation environment as shown in Fig. \ref{model}, where $M$ is the total number of snapshots in the dataset.
In addition,  loss function can be expressed as
\begin{equation}
	\label{math7}
	L = \frac{1}{M}\sum\limits_{m = 1}^M {\zeta (c_m^ \star ,\widehat {{c_m}})} ,
\end{equation}
where $\zeta ( \cdot )$ is the loss of a single snapshot, which measures the difference between   predicted value and  actual value.
Then, the goal is   to minimize the loss over all snapshots in the dataset $D$, which can be formally written as
\begin{equation}
	\label{math8}
	f_{{\Theta ^ \star }}^ \star  = \mathop {\arg \min }\limits_{{f_\Theta }} L(c_m^ \star |{{\rm Z}_m},\widehat {{c_m}}).
\end{equation}
The prediction function is parameterized by a set of model parameters $\Theta$ and is learned from the labeled data  in the dataset $D$. 
The objective is to find the best parameters $\Theta ^ \star$ that minimize the loss function. 
Next, we present proposed vision based solution for channel characteristic prediction.

\subsection{Framework}
\begin{figure*}[!t]
	\centering
	\includegraphics[width=.9\textwidth]{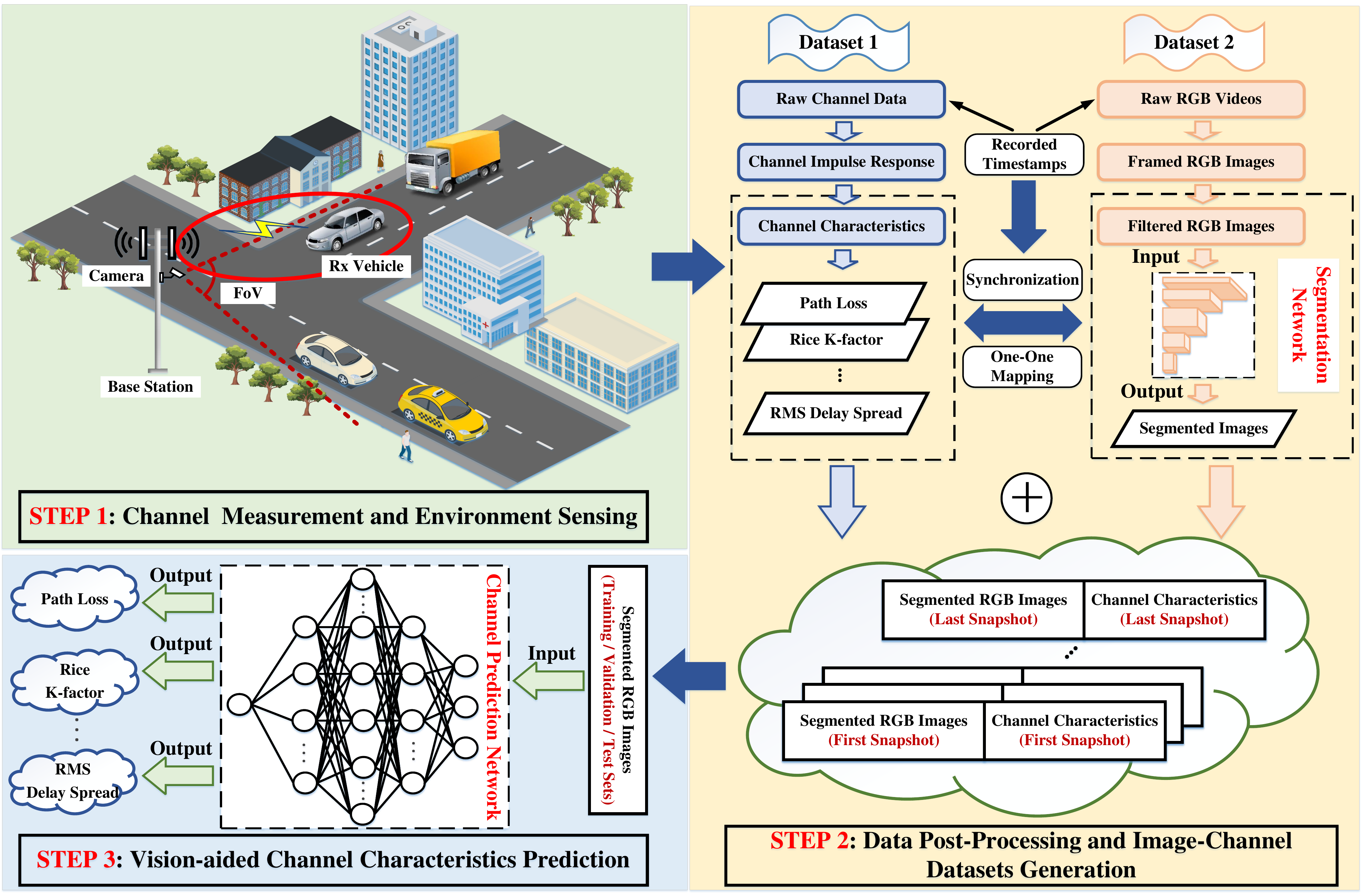}%
	\caption{Framework of proposed vision-based channel prediction model.}                    
	\label{model}
\end{figure*}

In this subsection, we propose a vision-aided deep learning model framework to predict channel propagation characteristics, with applicability extending beyond intersection scenarios to a wide range of V2I environments. 
The framework is built upon an image-channel dataset derived from measurement campaign, where  image dataset consists of RGB images of  target user captured at  BS, which are then processed into segmented images. 
By simply inputting segmented images into trained prediction network, accurate predictions of propagation characteristics, such as PL,  can be achieved.
The framework can be divided into three steps, as illustrated in Fig. \ref{model}, which will be introduced in detail below.

\textbf{Step 1: Channel Measurement and Environment Sensing.}
In Fig. \ref{model},  vector channel sounding signals are generated and transmitted by the BS, 
while the receiver is deployed on the moving target vehicle, 
thereby creating a typical dynamic  V2I scenario. 
As  channel sounding signals are continuously exchanged between  BS and  vehicle,   
RGB camera installed on the BS simultaneously captures videos with a wide-angle field of view (FoV), collecting environmental information surrounding the BS. 
The primary objective of this setup is to detect mobile user, specifically vehicle.
RGB camera  is selected due to small size, low cost, and ease of deployment. 
Furthermore,  RGB images captured are the most common form of visual data, consisting of three color channels, which are well-suited for accurately describing the real-world propagation environment. 
In Step 1, two distinct datasets are obtained: one comprising the raw channel data, typically in the form of in-phase/quadrature (I/Q) data, and the other consisting of the raw visual data, usually captured as video.

\textbf{Step 2: Data Post-Processing and Image-Channel Datasets Generation.}
To prepare the data for use in deep learning models, the raw channel and visual datasets must undergo a series of post-processing steps. 
For channel data, raw I/Q data is first converted into the channel impulse response. 
Subsequently, channel characteristic parameters are extracted through a set of parameter extraction algorithms, as outlined in Formulas (\ref{math2})--(\ref{math5}).
For raw visual data, RGB videos are processed frame by frame to generate individual RGB images. 
These images are then synchronized with channel impulse responses. 
Since snapshot acquisition rate of channel measurement system typically differs from frame rate of  RGB camera, a time-based alignment is necessary. 
The timestamps recorded during the acquisition of both  channel and visual data are compared, ensuring that each channel impulse response is accurately mapped to a corresponding RGB image. 
Any data that cannot be synchronized is discarded.
Additionally, a filtering step is applied to remove RGB images that are not suitable for further processing. 
Specifically, images where the mobile user is not visually identifiable are excluded. 
This can occur in the following situations: 
(1) camera lens is blocked by pedestrians or other vehicles, 
(2) mobile user is too far from the camera to be distinguishable, 
or (3) mobile user is obstructed by other vehicles or buildings. 
When such images are discarded, the corresponding channel data must also be removed to maintain data integrity.

Next, we also utilize a segmentation network to perform instance segmentation on  filtered images  to isolate  mobile user within each frame. 
This step is critical, as segmented  mobile user enhances the accuracy of subsequent channel prediction tasks. 
Detailed information can be found in  Section IV.
Upon completion of the post-processing, each segmented image is paired with corresponding channel characteristic parameters, such as PL and Rice K-factor. 
These pairs—comprising segmented RGB images and channel characteristics—form a unified image-channel  dataset suitable for training and validating deep learning models.

\textbf{Step 3: Vision-based Channel Characteristics Prediction.}
Channel characteristic prediction constitutes a regression problem within the realm of deep learning, necessitating a prediction network adept at handling such tasks.
The established  image-channel dataset is meticulously partitioned into training, validation, and test sets. 
Segmented RGB images serve as the input to  prediction network, while  channel characteristics function as the corresponding labels. 
Leveraging the training and validation sets, the prediction network delves into the inherent correlations and mappings between   images and channel characteristics. 
Through iterative updates, the network parameters are continually refined. 
Ultimately, the  performance is validated on test set. 
After training, the objective of  prediction network is to accurately forecast the channel propagation characteristics, including but not limited to PL, Rice K-factor,   RMS delay spread, by merely inputting segmented RGB images of mobile users in analogous scenarios.

\begin{figure*}[!t]
	\centering
	\includegraphics[width=.99\textwidth]{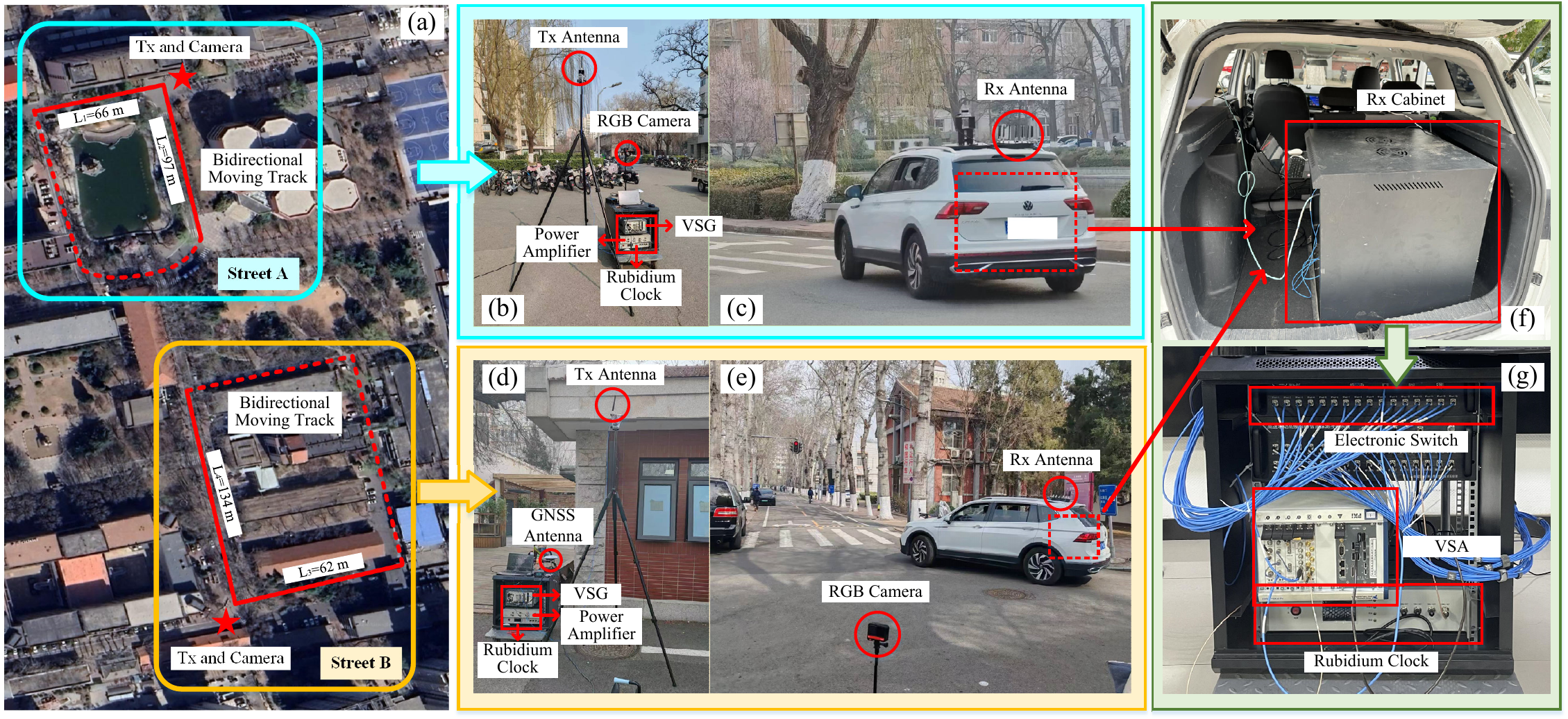}%
	\caption{Measurement scenarios for different streets.}                    
	\label{measurement}
\end{figure*}
  
\section{Experimental Setup}
\subsection{Datasets Generation}

The measurement  system, key equipment  and measurement scenarios  are depicted in Fig. \ref{measurement}. 
The measurement system consists of transmitting (Tx) and receiving (Rx) subsystems, with the core components being a vector signal generator (VSG) and a vector signal analyzer (VSA).
Specially, VSG and VSA   are   NI PXIe-5673 and NI PXIe-5663,  respectively. 
Tx and Rx antennas are vertically polarized omnidirectional antennas with identical specifications, covering a frequency range of 2–6 GHz. 
Tx side employs a single antenna, while Rx features a uniform circular array consisting of 16 individual antennas. 
These antennas are controlled by an electronic switch.
However, as the system model in this paper does not consider MIMO system  or angle-domain channel characteristics,  electronic switch is used solely to ensure the integrity of   RF line, and we don't use it to control and switch the sub-channels. 
Therefore, only the first antenna in  circular array is activated by default, effectively limiting the system to single-antenna channel data.
There are some other necessary auxiliary equipment.  
A power amplifier with a maximum gain of 30 dB is used to enhance the signal. 
For precise time synchronization, two rubidium clocks disciplined by Global Navigation Satellite System (GNSS) signals are employed. 
They also can provide real-time longitude and latitude coordinates, facilitating accurate positioning of the Tx and Rx.

\begin{table}[!t]
	\renewcommand{\arraystretch}{1.5}
	\begin{center}
		\caption{Configuration of Measurement System.} 
		\label{measure_system}
		\begin{tabular}{c c}
			\hline
			\hline
			Parameters              & Description \\  
			\hline
			Center Frequency               & 5.9 GHz \\
			Bandwidth               &      30 MHz \\
			Tx/Rx antennas     & \makecell[c]{Vertically polarized  \\ omni-directional antenna}  \\
			Transmitted power  & 30 dBm \\
			Snapshot acquisition rate  & 73 snapshots/second \\
			Resolution of original images & 1920$\times$1080  pixels\\
			Field of view of camera  & 120$^\circ$ \\
			Frame rate of camera     & 100 frames/second \\
			Height of Tx/Rx antennas and camera    & 3 m/2.1 m/1.8 m  \\
			Speed of vehicle & 15--20 km/h  \\
			\hline
			\hline
		\end{tabular}
	\end{center}
\end{table}

Most of  key equipment, excluding  antennas, is housed in two 12U cabinets, as illustrated in Figs. \ref{measurement}(b), (d)  and (f). 
The Rx cabinet is placed in the trunk of  test vehicle, with snapshot acquisition rate of 73 snapshots/second and the RF cable routed through a window to connect to Rx antenna, positioned approximately 2.1 meters high, as shown in Fig. \ref{measurement}(f). 
There is an inverter and an uninterruptible power supply  to provide continuous power supply.
Acting as a BS, Tx cabinet is placed and stationed at  intersection with Tx antenna mounted at a height of 3 meters.
The center frequency  is 5.9 GHz, which is a typical vehicular communication frequency band with a bandwidth of 30 MHz. 
The RGB camera, mounted near the transmitter at a height of 1.8 meters, has a FoV of 120$^\circ$ and a frame rate of 100 frames/second.
Rx vehicle which acting as a mobile user moves repeatedly in both directions along a predefined rectangular route at a speed of 15–20 km/h to ensure  sufficient visual data can be obtained. 
The visible portion of   routes, captured by  RGB camera, corresponds to the solid red lines in Fig. \ref{measurement}(a), while the red dotted lines represent areas outside the camera's field of view. 
The straight-line distance within the visible range varied from approximately 62 to 134 meters. 
Detailed parameters of   measurement system are shown in Table \ref{measure_system}.

Measurement campaign is conducted at two intersections in the urban area, as shown in Fig. \ref{measurement}(a). 
These two streets are located in the pedestrian zone, with dense pedestrian traffic, but relatively small vehicle traffic, and there is no situation where large vehicles completely block  Rx vehicle. Therefore,  channel  and visual data we obtain are based on the LoS  condition.

\subsection{Channel Data Post-processing}

Calibration of  measurement system is a critical process to ensure the accuracy and reliability of   experimental results. 
The process involves two primary components: back-to-back measurement and antenna calibration.   
Back-to-back measurement involves directly connecting   Tx and Rx subsystems using a  RF cable and a series of attenuators, bypassing the wireless channel. 
This setup compensates for amplitude-frequency response distortions introduced by system components such as cables, adapters, and transceivers.  
To address the impact of antenna characteristics, radiation patterns of both Tx and Rx antennas are meticulously measured in an anechoic chamber. 
This characterization facilitates the correction of errors arising from antenna radiation patterns, a procedure referred to as antenna calibration. 
These calibration steps establish a robust foundation for accurate measurement and analysis in wireless communication experiments.

Measurement system in this paper utilizes the frequency domain measurement method, 
and received signal can be described in  frequency domain as
\begin{equation}
	\label{math9}
	Y(f) = X(f){H_{Tx}}(f)H(f){H_{Rx}}(f),
\end{equation}
where $ X(f) $ and $ Y(f) $ are  transmitted  and received signal respectively, and $ H(f) $ is  channel transfer function.
Then $ {H_{Tx}}(f) $ and $ {H_{Rx}}(f) $ are  transfer functions of   Tx and Rx equipment  respectively, including cables, antennas, etc.
Received reference signal in back-to-back measurement which is to eliminate the influence of $ {H_{Tx}}(f) $  and $ {H_{Rx}}(f) $  can be expressed as
\begin{equation}
	\label{math10}
	{Y_{ref}}(f) = X(f){H_{Tx}}(f){H_{ref}}(f){H_{Rx}}(f),
\end{equation}
where $ {H_{ref}}(f) $ is transfer function of reference signal and is known.
Thus,  $ H(f) $   can be described as
\begin{equation}
	\label{math11}
	H(f) = \frac{{{Y_f}(f)}}{{{Y_{ref}}(f)}} \cdot {H_{ref}}(f).
\end{equation}
Further,  channel impulse response $ h(t,\tau ) $ can be obtained by inverse Fourier transform of   $ H(f) $.

\begin{figure}[!t]
	\centering
	
	\subfloat[]{\includegraphics[width=.42\textwidth]{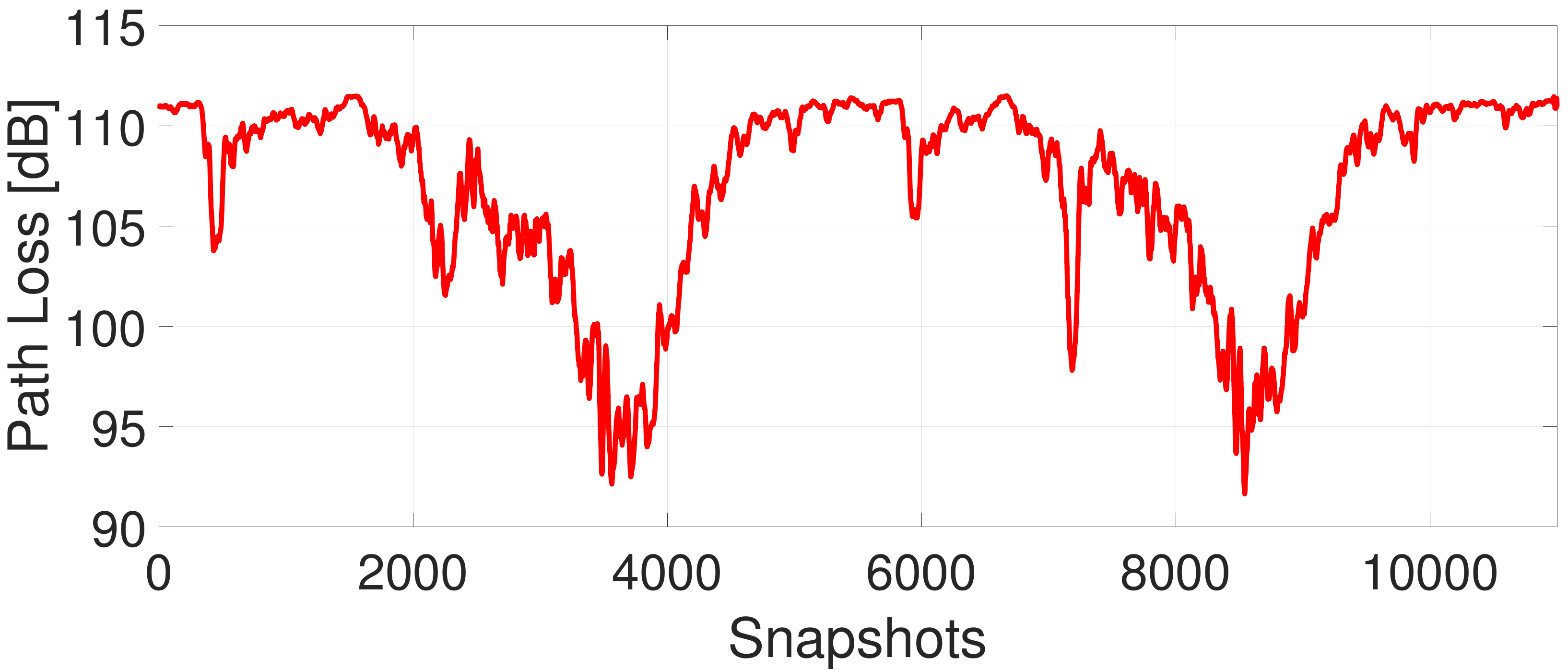}%
		\label{PL_eg}}
	\quad
	\subfloat[]{\includegraphics[width=.42\textwidth]{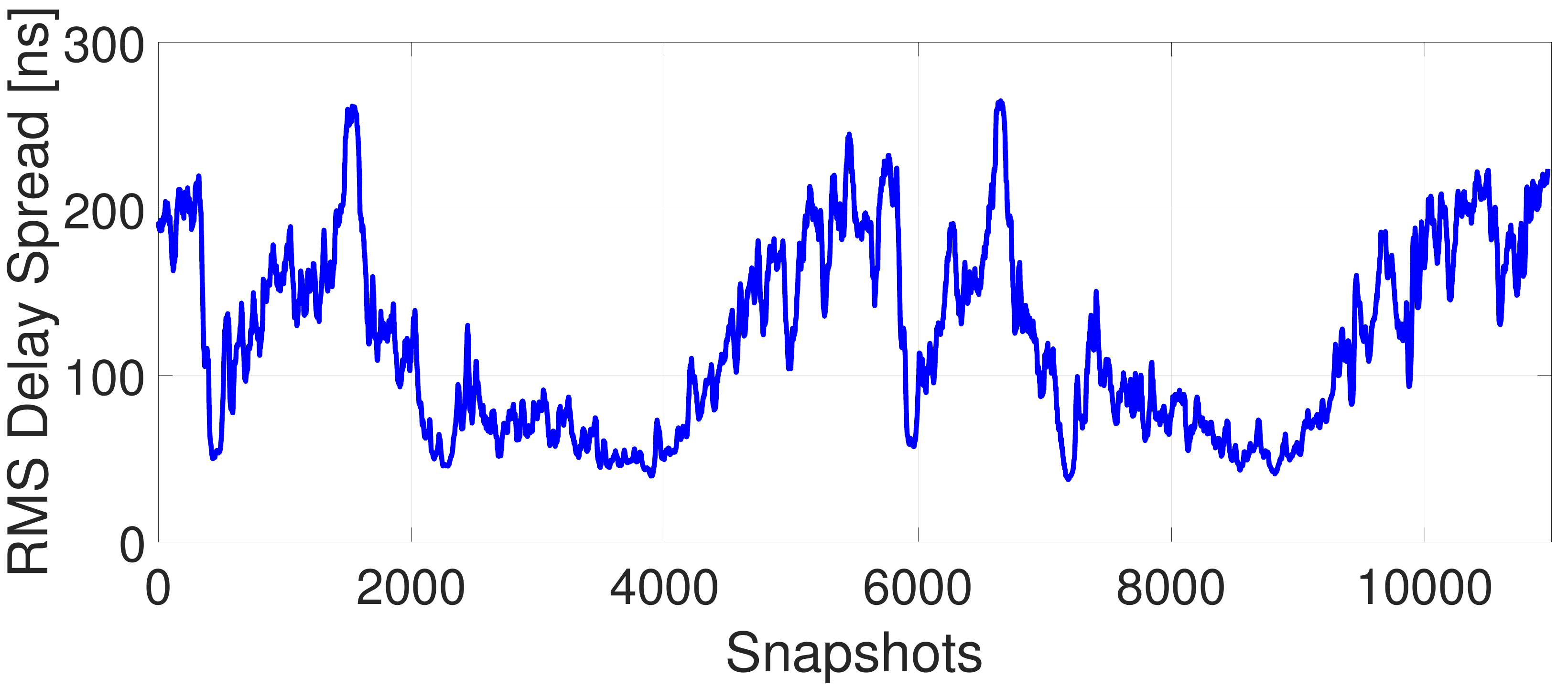}%
		\label{DS_eg}}
	\quad
	\subfloat[]{\includegraphics[width=.42\textwidth]{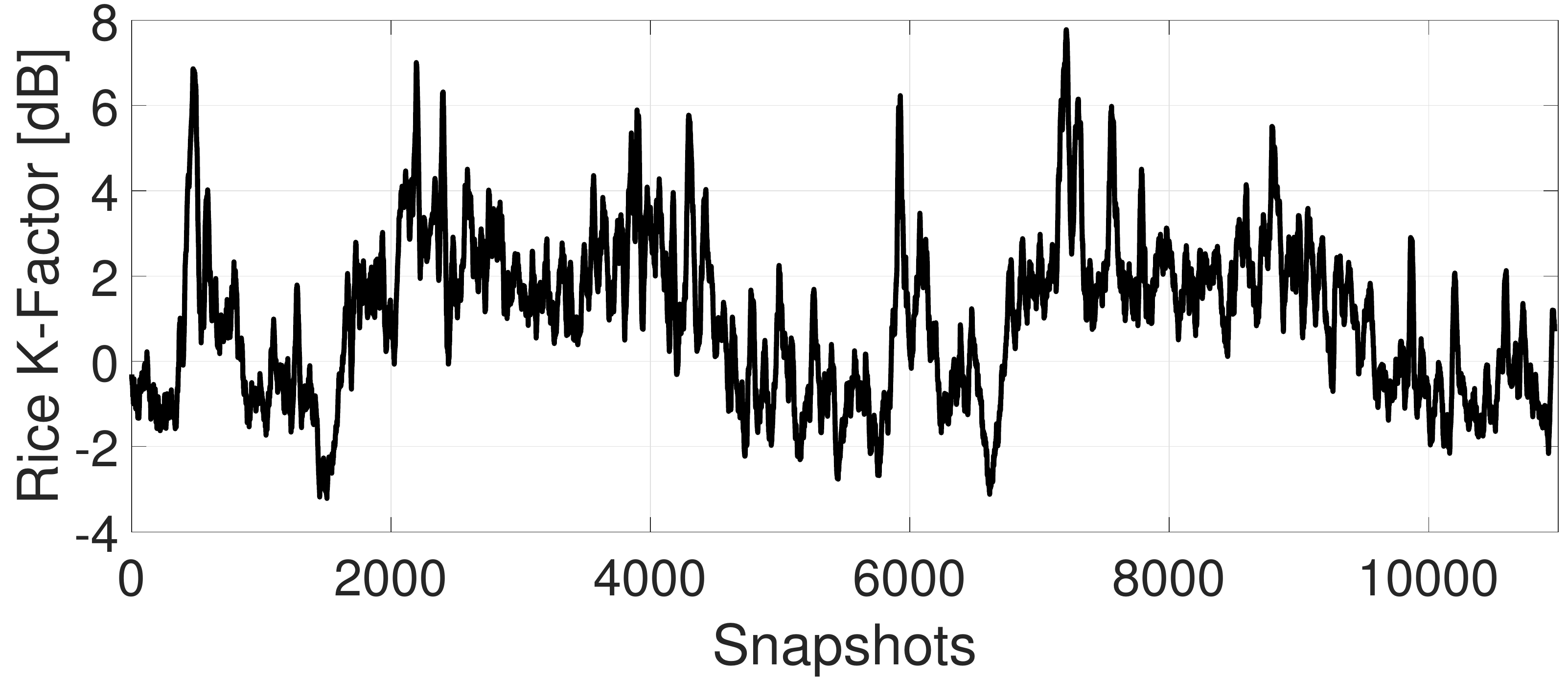}%
		\label{K_eg}}
	\quad
	\subfloat[]{\includegraphics[width=.42\textwidth]{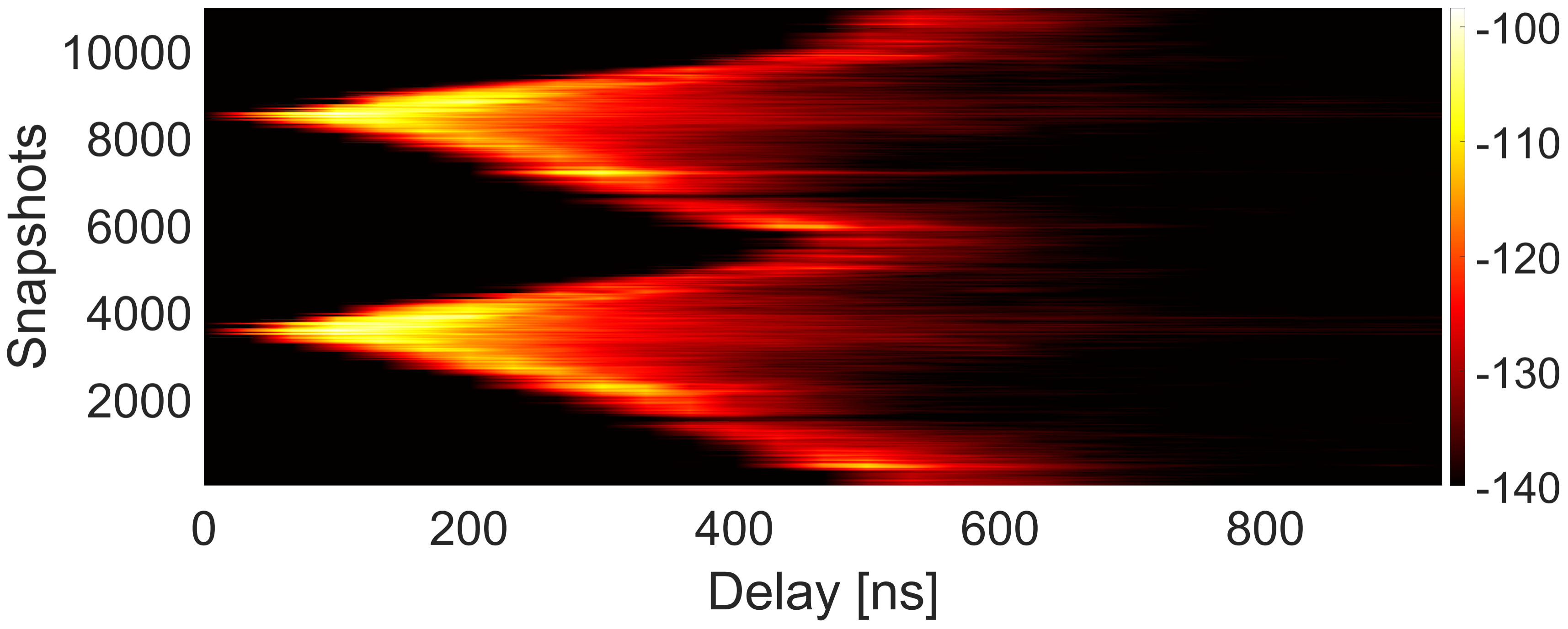}%
		\label{PDP_eg}}
	\caption{
		Examples of measurement-based  channel characteristics.
		(a) PL;
		(b) RMS delay spread;
		(c) Rice K-factor;
		(d) APDP.
	}
	\label{eg}
\end{figure}

Using  channel parameter extraction algorithms, as defined in Formulas (\ref{math2})--(\ref{math5}), a series of channel propagation characteristics are obtained. 
For illustrative purposes, measurement data from two complete laps along Street A   in Fig. \ref{measurement}(a) is analyzed. 
Key channel characteristics, including PL, PDP, RMS delay spread, and Rice K-factor, are presented in Fig. \ref{eg}. 
Notably, the  four parameters  are plotted as functions of snapshot index, which correlates with  measurement time.  
On the one hand, it allows for a direct comparison with the real-time measurement conditions, and on the other hand, it is appropriate given the limited transceiver distance range, which does not exceed 150 meters. 
Although PL and K-factor are conventionally modeled as distance-dependent, the snapshot-based dimension is more suitable for this paper.  
The results indicate a clear physical relationship between the temporal variation of these parameters and the movement of  Rx vehicle relative to BS. 
Specifically, as the Rx vehicle approaches and subsequently moves away from   BS,
PL and RMS delay spread first decrease and then increase, while Rice K factor first increases and then decreases.  
These observed trends align with the expected behavior of  channel under  given measurement conditions, demonstrating the validity and effectiveness of both   measurement system and   parameter extraction algorithms.

\subsection{Visual Data Post-processing}
The original visual data is captured as RGB video, from which individual RGB frames are extracted at a rate of 100 frames/second, yielding images with a resolution of 1920 × 1080 pixels. 
As mentioned in Section II-C, there is a mismatch between  sampling rates of  visual and channel data. 
Thus a timestamp-based alignment method, which is commonly employed in dataset construction in \cite{deepsense, imran2024, Charan2021}, is used to synchronize  the two datasets. 
After synchronization, redundant visual data frames are removed to ensure consistency.
Furthermore, an additional filtering process is applied to  synchronized image data. 
Specifically, images that either failed to capture the Rx vehicle or where the vehicle was indistinguishable upon visual inspection are excluded. 
Correspondingly, the associated channel data for these excluded images is also discarded. 

In intersection scenarios, characterized by dense vehicular and pedestrian traffic, propagation environment contains dense dynamic scatterers. 
Directly feeding the filtered images into  prediction network poses a significant challenge, as the network may struggle to extract key channel-related information from complex images saturated with redundant data. 
To address this issue and enhance the learning capacity of  prediction network, we propose performing instance segmentation specifically for mobile user, i.e.  Rx vehicle.
Segmented images emphasize critical information, including the shape, size, and spatial position of   Rx vehicle. 
Additionally,  segmentation shapes provide insights into  driving direction, enabling determination of whether   Rx vehicle is moving counterclockwise, clockwise, going straight, or turning.
The aforementioned CV-based method has also been employed in Refs. \cite{env,feifei2023,feifeigao2023,imran2024} to assist with tasks such as mmWave beam prediction and blockage prediction. 
Environmental information conveyed by  segmented images is referred to as  environmental semantics  in these studies, highlighting its role in capturing critical contextual features relevant to channel propagation.

The You Only Look Once (YOLO) family of algorithms is a widely recognized deep learning framework frequently employed in CV applications. 
Among its iterations, YOLOv8, released by Ultralytics in 2023 \cite{Jocher_Ultralytics_YOLO_2023}, has demonstrated substantial improvements in both accuracy and speed compared to its predecessors.  
The architecture of YOLOv8 mainly consists of three parts: Backbone, Neck and Head. 
Backbone extracts multi-scale features from  input images, while Neck performs feature fusion across multiple scales, and finally Head simultaneously generates target detection boxes and instance segmentation masks. 
This efficient design makes YOLOv8 particularly well-suited for scenarios with stringent real-time performance requirements.

\begin{figure}[!t]
	\centering
	\includegraphics[width=.5\textwidth]{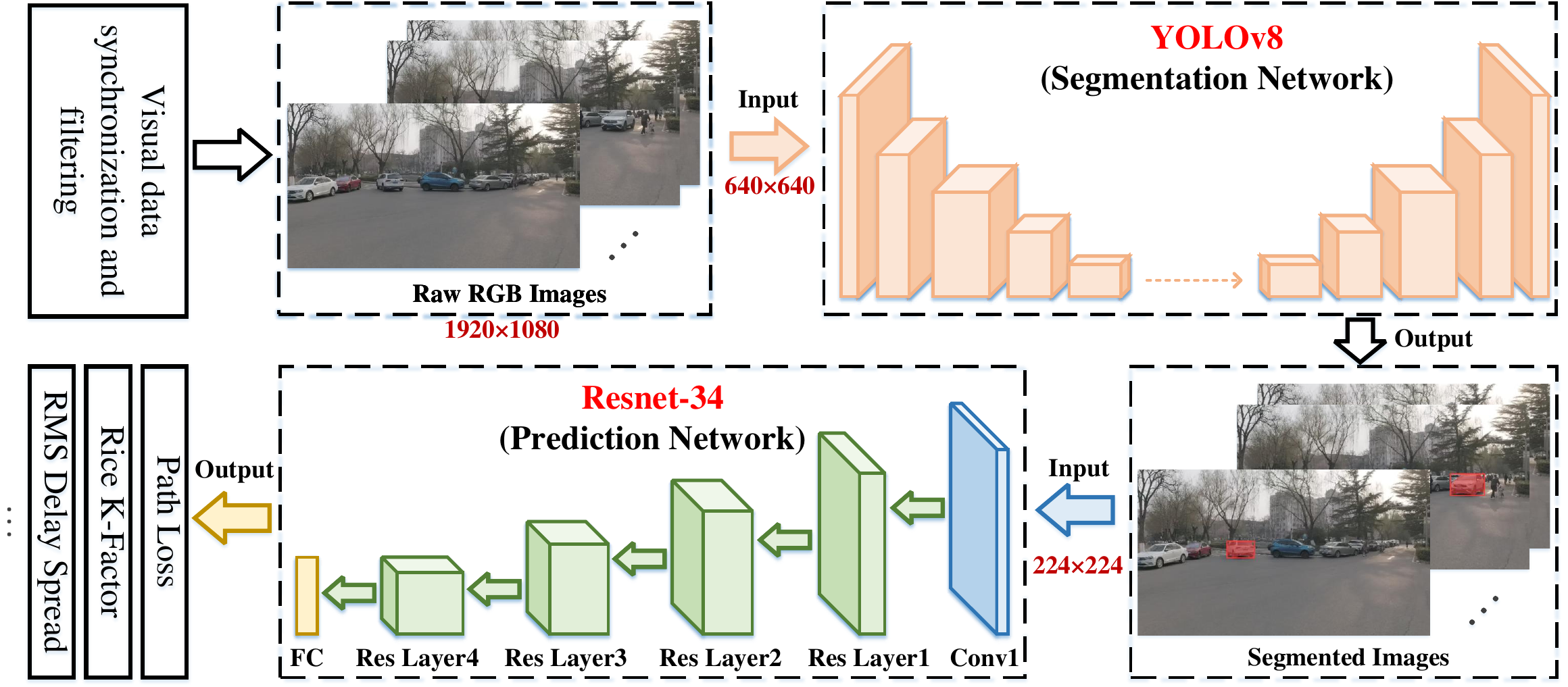}%
	\caption{Workflow diagram of image segmentation network and channel prediction network.}                    
	\label{network}
\end{figure}

Ultralytics also provides pre-trained YOLOv8 models, which are trained on the COCO dataset \cite{COCO}. These pre-trained models greatly facilitate the extraction of environmental information through transfer learning, making YOLOv8 an ideal segmentation network for our study.  
It should be noted, however, that while the segmentation accuracy of  pre-trained YOLOv8 model is high, certain challenges arise in intersection scenarios. 
Rx Vehicle  may appear in the images from varying angles, as illustrated in Fig. \ref{network}, and may be partially occluded by pedestrians or other vehicles. 
These factors can degrade the segmentation accuracy of YOLOv8.  
To address this issue, we adopt an iterative training approach. 
Specifically, images with noticeably poor segmentation results from   pre-trained YOLOv8 model are extracted and manually annotated to form a new dataset. 
This dataset is then used to fine-tune  YOLOv8 network iteratively. 
Through this process, we achieve consistently high segmentation accuracy for Rx vehicles across all images.

\begin{table}[!t]
	\renewcommand{\arraystretch}{1.2}
	\begin{center}
		\caption{Configuration of Hyperparameters.} \label{table2}
		\begin{tabular}{c c c}
			\hline
			\hline
			Hyperparameters          & YOLOv8  & ResNet-34 \\   
			\hline
			Batch size              & 16      & 64  \\
			Epochs                  & 500     & 60  \\
			Learning rate           & 0.01    & 0.001 \\
			Input image size        & 640$\times$640  pixels & 1920$\times$1080  pixels \\
			Loss function  & Binary Cross-Entropy  &  Mean Square Error  \\  
			Optimizer         &  AdamW   &  Adam \\
			Activation function   & Sigmoid    &  ReLu \\
			\hline
			\hline
		\end{tabular}
	\end{center}
\end{table}

\subsection{Dataset and Model Training}
After completing   data post-processing, we obtain the image-channel dataset  $D = \left\{ {\left( {{Z_m},c_m^ \star } \right)} \right\}_m^M$, where $Z_m $ and $c_m^ \star$ are  the segmented image  and the value of   channel characteristic of the  $m$th snapshot, respectively.
What's more, $c_m^ \star \in \Omega $,  $\Omega  = [PL,K,{\tau _{rms}}]$.
Dataset $D$ comprises four sub-datasets: one segmented image dataset and three channel characteristic datasets. 
These sub-datasets share the same number of snapshots, ensuring that \( M \) remains consistent across all four components.  
Specifically, for measurements conducted on Street A, 11 laps are completed, resulting in 11652 valid  snapshots. 
Similarly, 9 laps are conducted and  15045 valid  snapshots are obtained on Street B. 

For model training, both segmentation network  and  prediction network are trained. 
The training process for YOLOv8 has been detailed in Section III-C, and its hyperparameter settings align with those described in  \cite{Jocher_Ultralytics_YOLO_2023}. 
No modifications are made to its network architecture.  
As for prediction network, it is required robust image feature extraction capabilities and must effectively handle regression tasks, and we adopt and customize  ResNet-34  proposed in \cite{he2015deep} to address   channel characteristic prediction problem, which is also used in  \cite{AhmedAlkhateeb2022, env}.  
The structural diagram of ResNet-34, presented in Fig. \ref{network}, can be broadly divided into three components:  input layer,  residual  layers, and  fully connected (FC) layer. 
The input layer comprises a $7 \times 7$ convolutional layer followed by a $3 \times 3$ max pooling layer, with an input size of $224 \times 224$. 
Residual layers are organized into 4 stages, each containing multiple residual blocks, where each block consists of two $3 \times 3$ convolutional layers. 
Output of  final residual layer is fed into a global average pooling layer, which compresses   spatial dimensions, followed by a fully connected layer that produces the predicted values of  channel characteristics.

\begin{figure*}[!t]
	\centering
	\includegraphics[width=.90\textwidth]{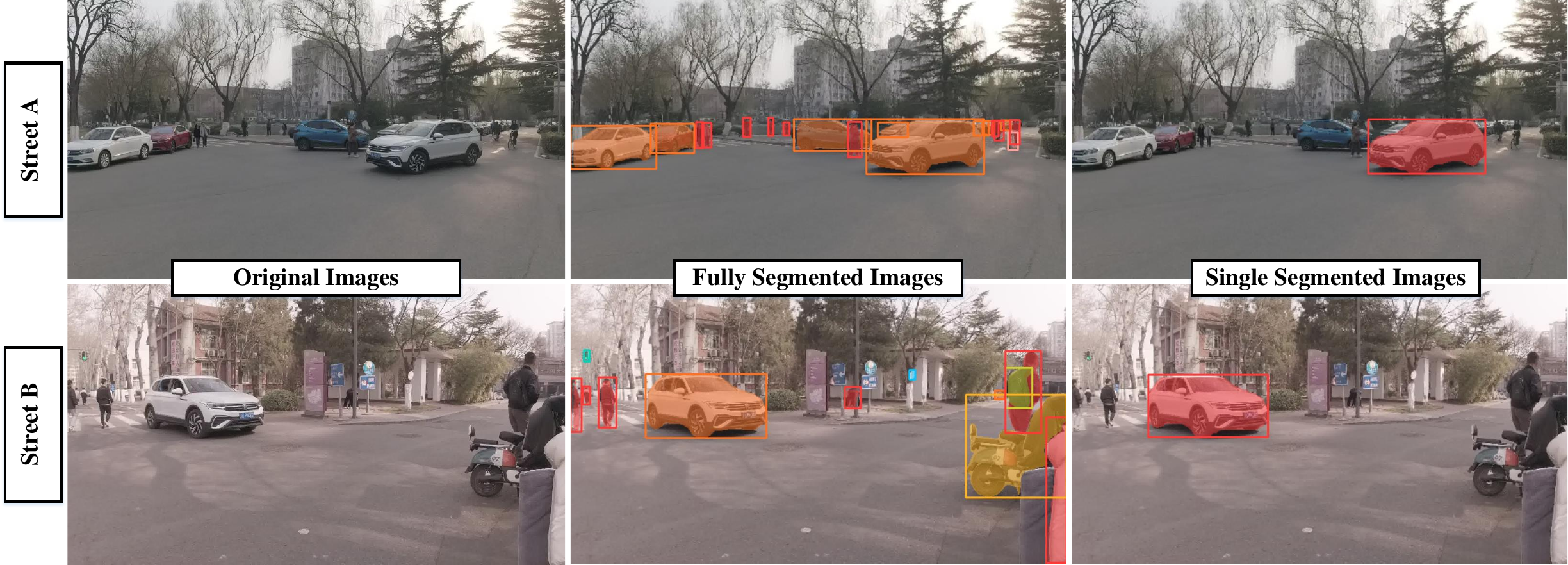}%
	\caption{Examples of three types of images on Streets A and B.}                    
	\label{images}
\end{figure*}

Notely, network structure of ResNet-34  utilized in this paper remains consistent with \cite{he2015deep}, while the original fully connected layer is replaced with a single fully connected layer that produces one output. 
This modification ensures compatibility with   regression nature of   channel characteristic prediction task, where the final output corresponds to specific channel parameters such as PL  and Rician K-factor (in dB) or RMS delay spread (in ns).  
The customized ResNet-34  is then fine-tuned in a supervised manner using labeled segmented images, where each image is paired with its corresponding channel characteristics. 
This fine-tuning process ensures that ResNet-34 is optimized for accurately predicting channel characteristics in the given environment.  

Key  hyperparameters   of model training are shown in  Table \ref{table2}. 
The  architecture of YOLOv8 is relatively complex. 
More detailed explanation of its structure and hyperparameters can be found in \cite{Jocher_Ultralytics_YOLO_2023}.
For ResNet-34, number of training epochs  and batch size are 60 and 64 respectively, with learning rate   0.001. 
Adam \cite{kingma2014adam} is selected as optimization algorithm   and  mean square error   function is selected as  loss function, with activation function of ReLU.
Both YOLOv8 and ResNets are trained using  deep learning framework PyTorch.

\section{Performance evaluation and analysis}
Based on   image-channel datasets we built, we carry out a a series of experiments to fully evaluate and analyze the performance of  proposed   model on channel characteristics prediction, including PL, Rice K-factor, and RMS delay spread. 
Root mean square error (RMSE) is chosen as   metric to evaluate  prediction accuracy of model, which is given by
\begin{equation}
	\label{math12}
	RMSE = \sqrt {\frac{1}{M}\sum\limits_{i = 1}^M {{{({y_i} - \widehat {{y_i}})}^2}} },
\end{equation}
where $\mathbf{y} = \left\{ {{y_1},...,{y_M}} \right\}$ and $\mathbf{\widehat y} = \left\{ {\widehat {{y_1}},...,\widehat {{y_M}}} \right\}$ refer to the measured and predicted values, respectively.

Moreover, to investigate the impact of separately segmenting   images of Rx vehicle  on improving  prediction accuracy, two additional sets of comparative images are prepared. 
They are  original RGB images and   fully segmented images as shown in Fig. \ref{images}.
As for fully segmented images, all common objects, such as pedestrians, bicycles and non-test vehicles,  are segmented and assigned a uniform color mask and bounding box to objects belonging to the same category.

\begin{figure}[!t]
	\centering
	
	\subfloat[]{\includegraphics[width=.45\textwidth]{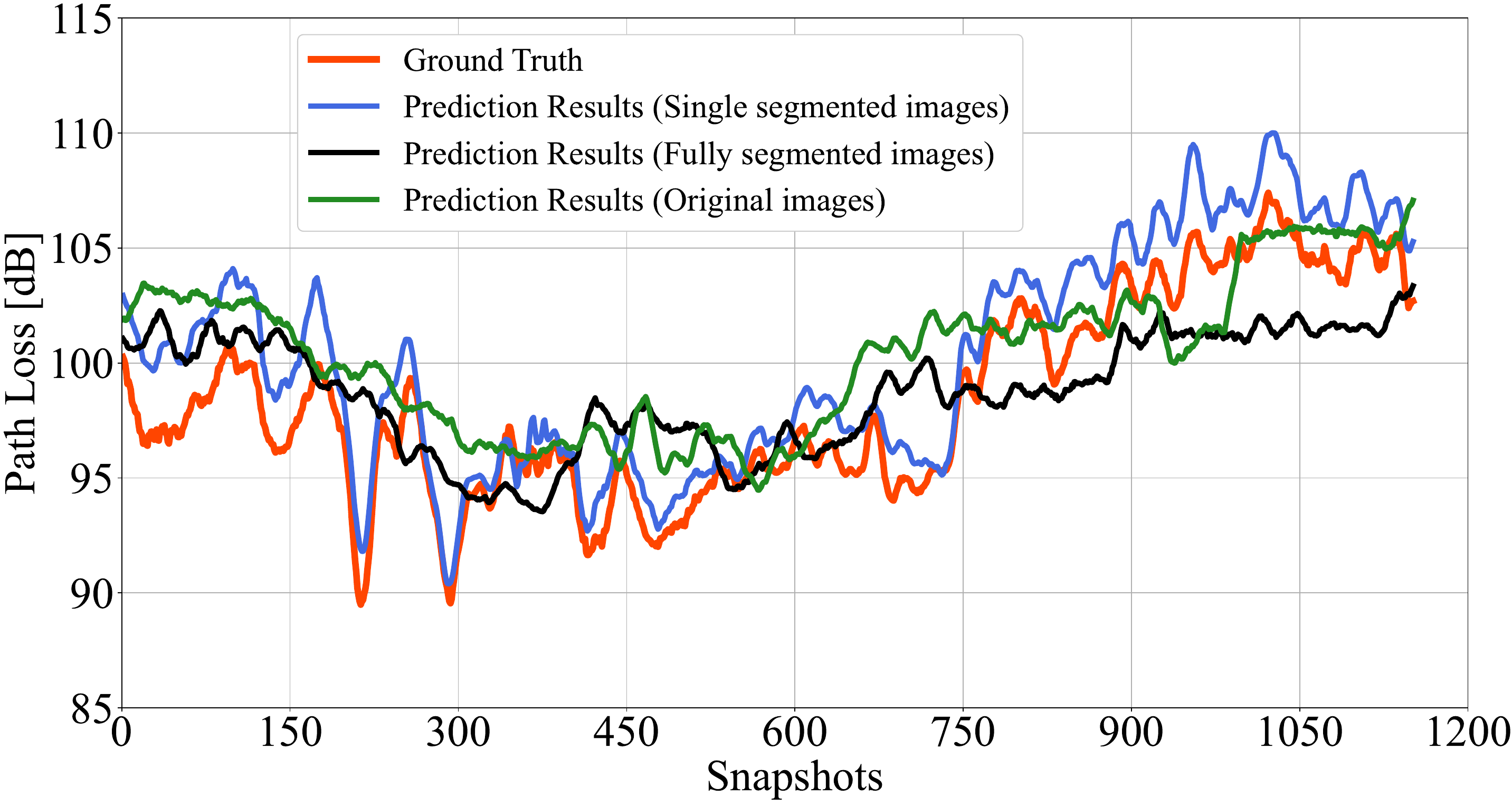}%
		\label{BaJiao_Day_1_PL_Exp1}}
	\quad
	\subfloat[]{\includegraphics[width=.45\textwidth]{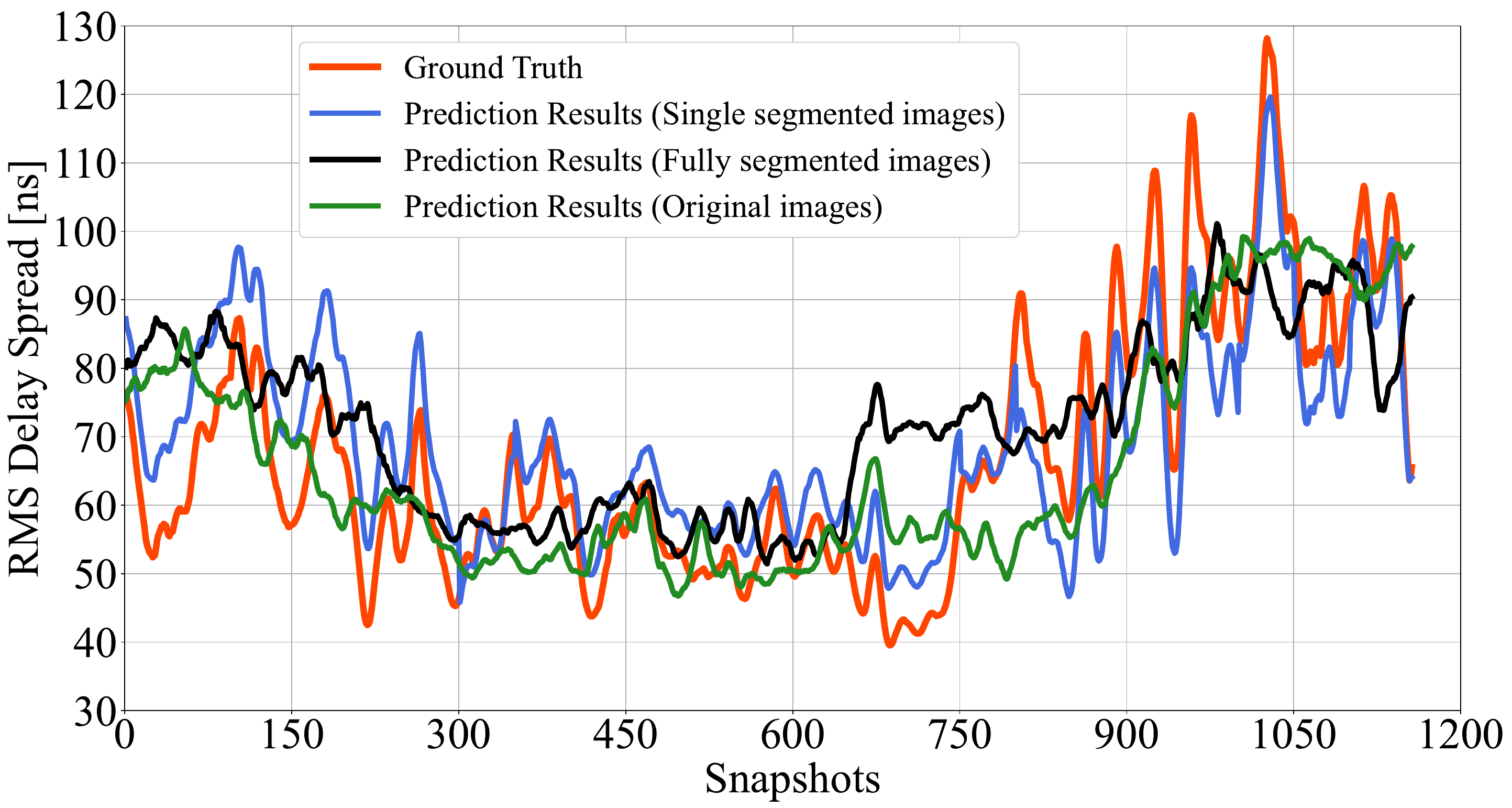}%
		\label{BaJiao_Day_1_DS_Exp1}}
	\quad
	\subfloat[]{\includegraphics[width=.45\textwidth]{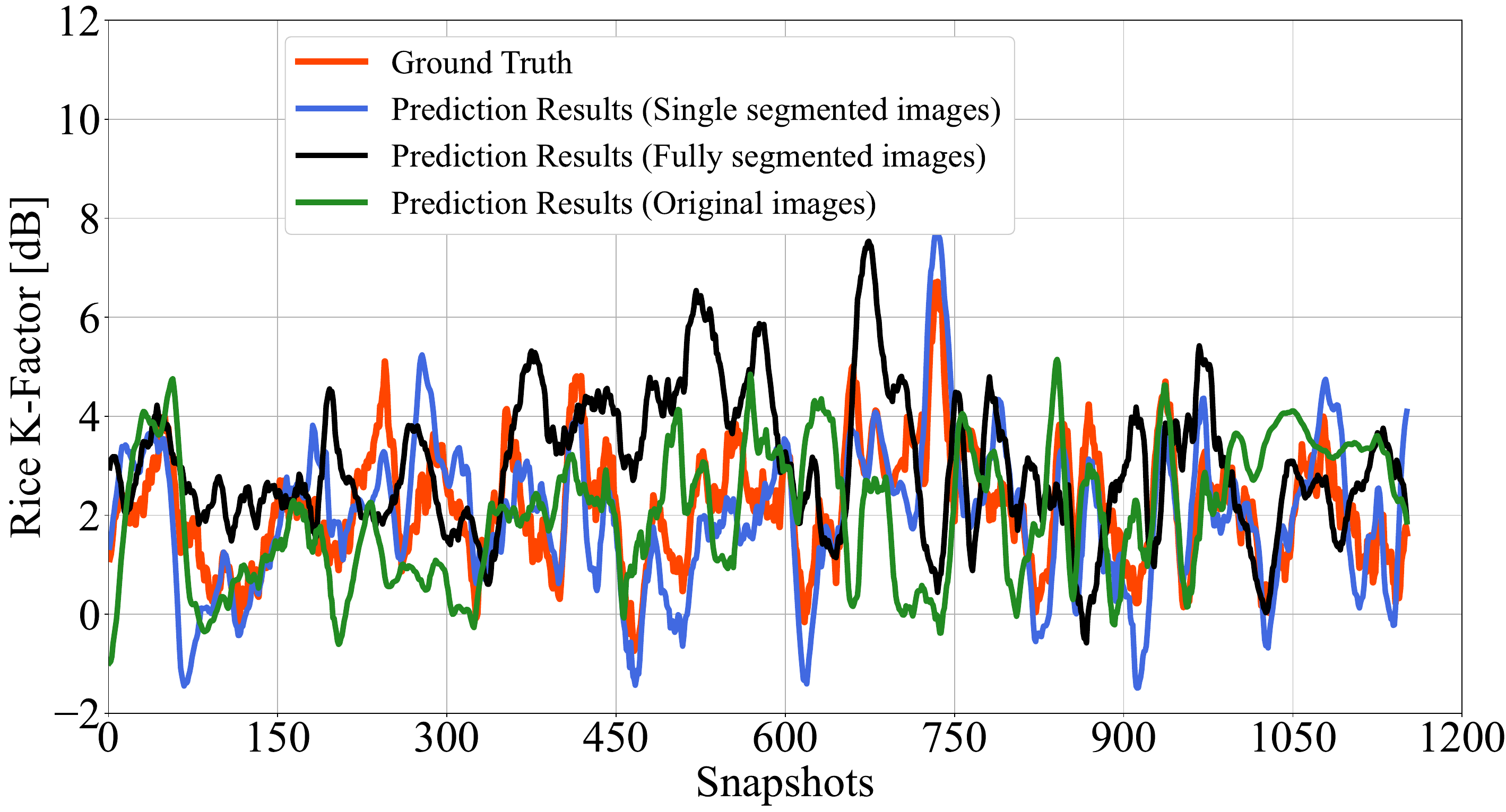}%
		\label{BaJiao_Day_1_K_Exp1}}
	\quad
	\caption{
		Prediction results of three channel characteristics in Experiment 1 on Street A.
		(a) PL;
		(b) Rice K-factor;
		(c) RMS delay spread.
}
	\label{exp1}
\end{figure}

\subsection{Experiment 1: Scenario Self-Validation }

\begin{table*}
	\renewcommand{\arraystretch}{1.5} 
	\begin{center}
		\caption{RMSE comparison of prediction accuracy in different scenarios.}
		\label{RMSE}
		\begin{threeparttable}
			\begin{tabular}{ c| c| c| c| c| c}
				\hline
				\hline
				Experiments &   \makecell[c]{Test Sets and Models} & \makecell[c]{Image Types} & \makecell[c]{Path Loss \\\ [dB]} &  \makecell[c]{Rice K-Factor \\\ [dB]}  & \makecell[c]{RMS Delay Spread \\\ [ns]}    \\
				\hline
				\makecell[c]{\makecell[c]{Scenario \\ Self-Validation}}    & \makecell[c]{\makecell[c]{Dataset A $\rightarrow$ Model A /  Dataset B $\rightarrow$ Model B}}  & \makecell[c]{Original \\ Fully segmented \\ Single segmented} & \makecell[c]{4.86 / 4.61 \\ 3.09 / 3.92 \\ \textbf{2.46} / \textbf{2.63}} & \makecell[c]{5.67 / 4.41 \\ 4.92 / 4.32 \\ \textbf{2.37} / \textbf{2.45}} & \makecell[c]{11.61 / 13.55 \\ 14.23 / 12.92\\ \textbf{2.70} / \textbf{2.91}} \\
				\hline
				\makecell[c]{\makecell[c]{Scenario \\ Cross-Validation}}    & \makecell[c]{\makecell[c]{Dataset A $\rightarrow$ Model B /  Dataset B $\rightarrow$ Model A}}  & \makecell[c]{Original \\ Fully segmented \\ Single segmented} 
				& \makecell[c]{8.68 / 9.16 \\ 9.66 / 9.39 \\ \textbf{4.9} / \textbf{4.68}} 
				& \makecell[c]{7.54 / 8.52 \\ 8.45 / 7.92 \\ \textbf{3.81} / \textbf{4.56}} 
				& \makecell[c]{42.74 / 33.25 \\ 25.69 / 22.47\\ \textbf{6.22} / \textbf{5.95}} \\
				\hline
				\makecell[c]{\makecell[c]{Non-test Vehicle \\ Cross-Validation}}    & \makecell[c]{\makecell[c]{Test set of other vehicle A $\rightarrow$ Model A\\  
						/ Test set of other vehicle B $\rightarrow$ Model B}}  & \makecell[c]{Single segmented} 
				& \makecell[c]{\textbf{2.94} / \textbf{2.87} } 
				& \makecell[c]{\textbf{3.27} / \textbf{2.98} } 
				& \makecell[c]{\textbf{3.18} / \textbf{3.45} } \\
				\hline
				\hline
			\end{tabular}
		\end{threeparttable} 
	\end{center}
\end{table*}

The datasets from Streets A and   B, referred to as Dataset A and Dataset B, are divided into training, validation, and test sets in a ratio of 0.8, 0.1, and 0.1, respectively. 
To ensure temporal continuity and capture  dynamic characteristics in V2I scenarios, test sets consist of continuous snapshots, while  training   and validation sets are randomly assigned after removing  test sets.
Subsequently,  ResNet-34 network is trained and evaluated on Datasets A and   B, resulting in models designated as Model A and Model B, respectively. 
Note that three types of images in Fig. \ref{images} are used to train  models separately. 
In other words, Models A and B actually contain three sub-models respectively. 
The input of  sub-models during training is the three types of images, and the labels are the same channel characteristics  values.
The prediction accuracy of these  models on  PL, Rice K-factor, and RMS delay spread is presented in Table 3. 
Taking Street A as an example, the prediction results are shown in Fig. \ref{exp1}.

When single segmented images are used as input, the predicted results align closely with  actual variations of  three channel characteristics As illustrated in Fig. \ref{exp1}. 
In contrast, the predictions derived from the other two types of input images fail to capture detailed dynamic fluctuations, instead only approximating the general upward or downward trends. 
This observation is further supported by the results presented in Table \ref{RMSE}. 
Specifically, when a single segmented image serves as the input, the prediction accuracy for   three channel parameters achieves deviations of less than 3 dB or 3 ns. 
However, with the other two image types as inputs, the prediction errors for PL,  Rice K-factor, and RMS delay spread fall within the ranges of 3.09–4.86 dB, 4.32–5.67 dB, and 11.61–14.23 ns, respectively. 
The prediction error for RMS delay spread is particularly pronounced.
Moreover, within the dataset collected from the same street, the overall background across all images exhibits minimal variation. 
While dynamic objects such as pedestrians and other vehicles may sporadically appear, these occurrences are random and short-lived, typically spanning only a few dozen frames. 
As the analysis is confined to LoS condition, Rx vehicle remains consistently visible in all images without being entirely occluded. 
Consequently, pixel variations between consecutive images primarily result from changes in the position and size of Rx vehicle. 
This also leads to the ability to obtain fuzzy but inaccurate channel change trends when inputting original images and fully segmented images.

 \begin{figure}[!t]
	\centering
	
	\subfloat[]{\includegraphics[width=.45\textwidth]{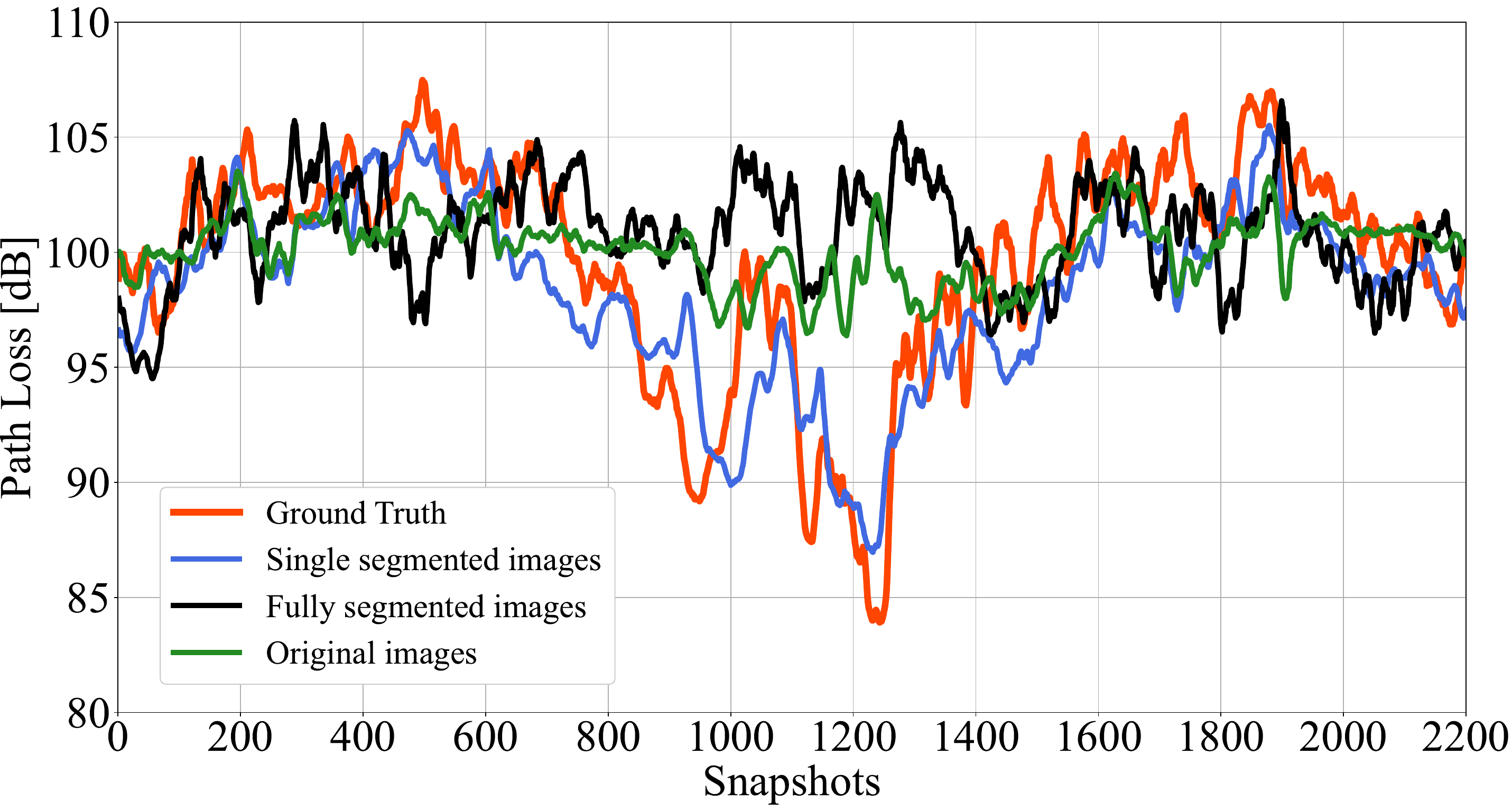}%
		\label{BaJiao_Day_1_PL_Exp2}}
	\quad
	\subfloat[]{\includegraphics[width=.45\textwidth]{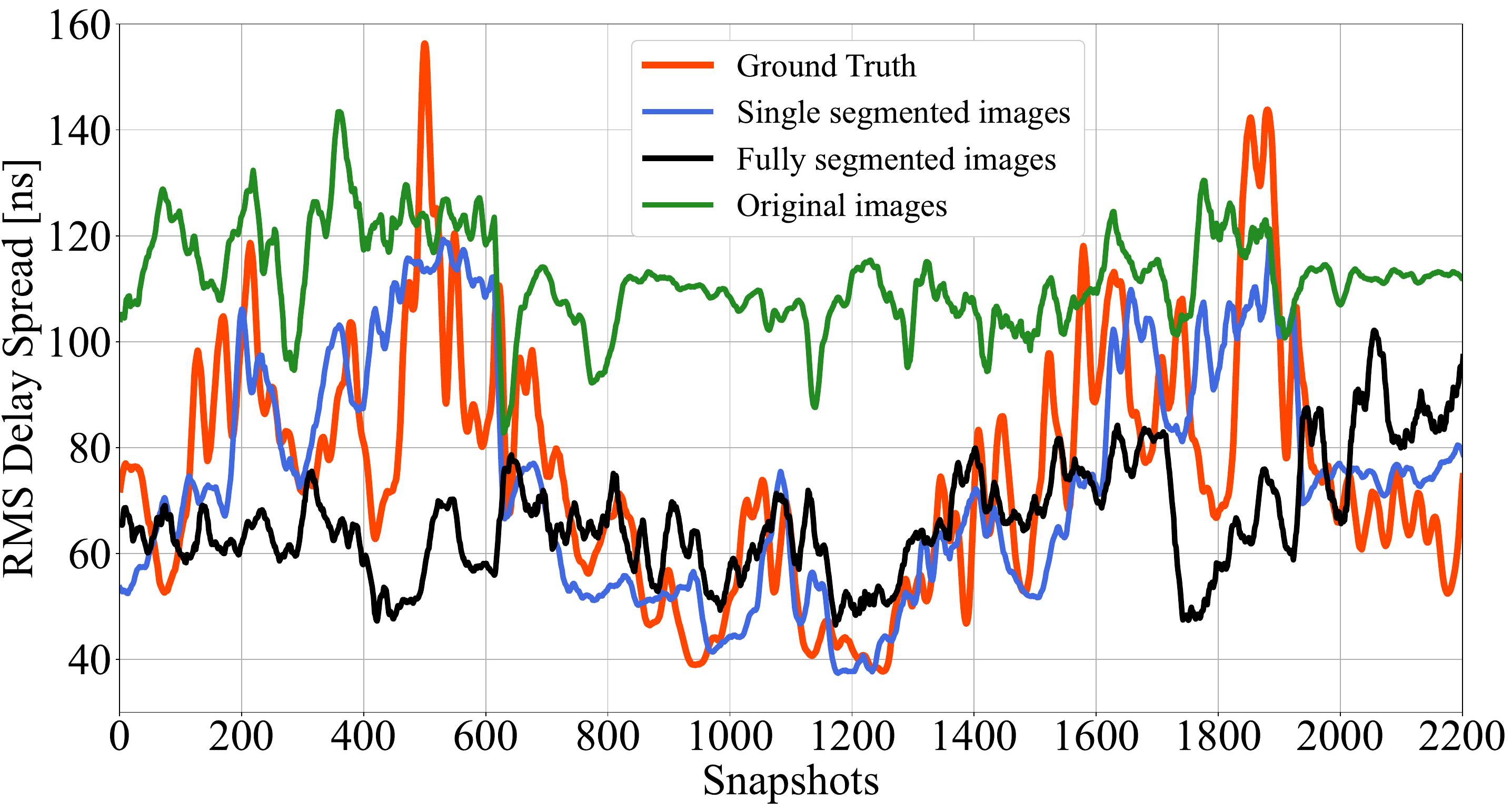}%
		\label{BaJiao_Day_1_DS_Exp2}}
	\quad
	\subfloat[]{\includegraphics[width=.45\textwidth]{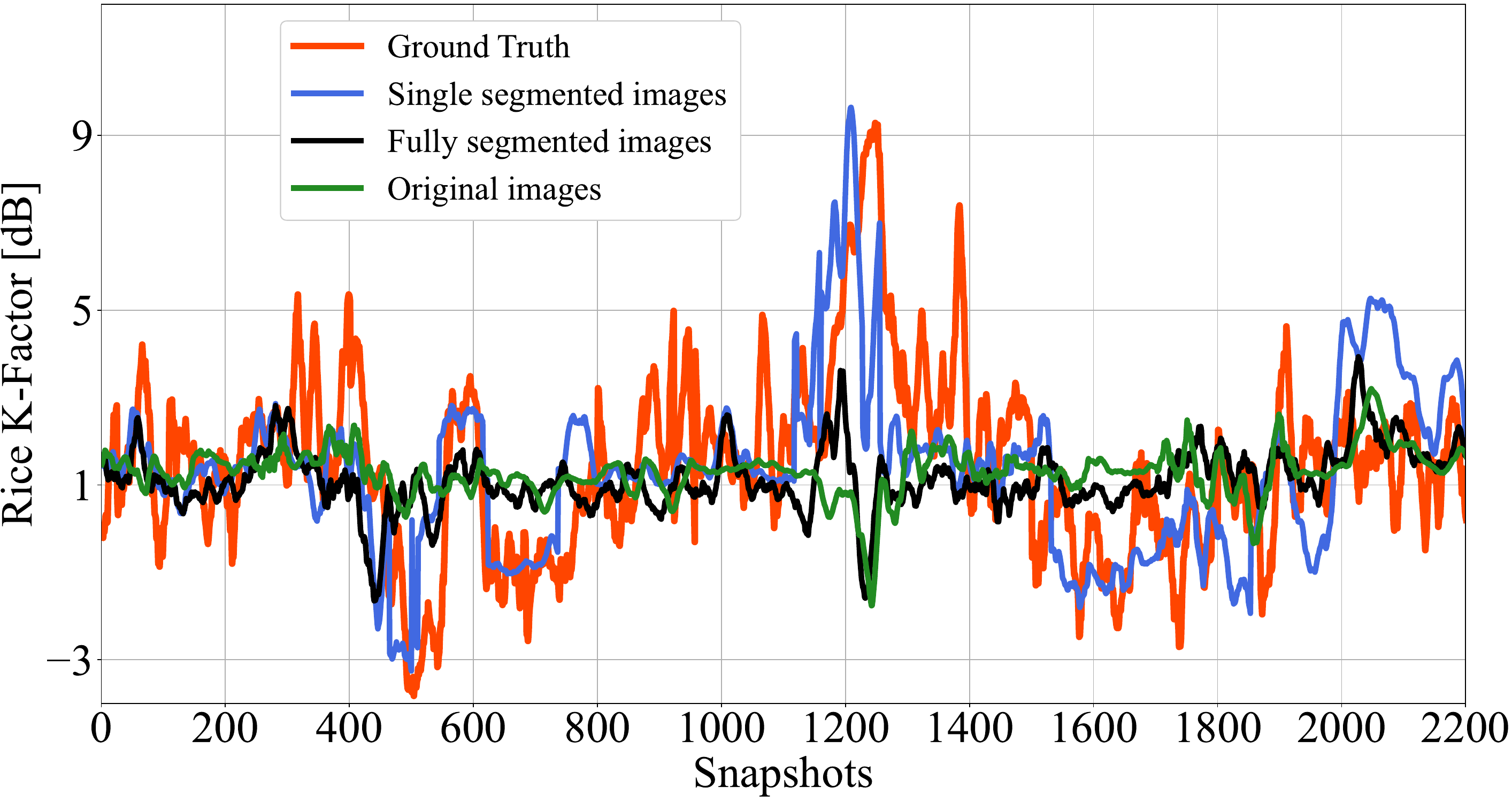}%
		\label{BaJiao_Day_1_K_Exp2}}
	\quad
	\caption{
		Prediction results of three kinds of images in Experiment 1.
		Case A of Street 1: (a) BBox;
		(b) Instance Segmentation;
		(c) Binary Mask.
	}
	\label{exp2}
\end{figure}

\subsection{Experiment 2: Scenario Cross-Validation }
In this subsection, we evaluate the generalization capability of   models. 
Models A and B, which are trained in Section IV-A, are now tested using data from both Dataset A and Dataset B. 
Specifically, we use approximately 2200 consecutive data snapshots from each dataset as test sets. 
To assess model performance, we perform cross-validation by inputting the test set from Dataset A into Model B, and the test set from Dataset B into Model A. 
Similar to Section IV-A, we compare the results using both  original images and  fully segmented images as comparison. 
The RMSE results of cross-validation are listed in Table \ref{RMSE}.
Taking Model A as an example, the prediction results of three channel characteristics after inputting   test set of Dataset B are shown in  Fig. \ref{exp2}.

As shown in Table \ref{RMSE}, the prediction errors of all models increased in the cross-validation experiment. 
Specifically, RMSEs for the models trained and tested with  single segmented images when predicting PL, Rice K-factor, and RMS delay spread range from 3.81-4.56 dB, 4.68-4.90 dB, and 5.95-6.22 ns, respectively. 
In contrast,  prediction errors of the other two images as data sets for the above parameters increased significantly, with RMSEs of 8.68-9.66 dB, 7.54-8.52 dB, and 22.47-42.74 ns, respectively. Compared to Experiment 1, RMSEs for the latter two images show a more substantial increase, indicating a sharper decline in prediction performance.
Consistent with  Experiment 1,  RMSEs for RMS delay spread prediction remain  the largest. 
It is closely related to the fact that RMS delay spread has a larger range in the datasets, with fluctuation within 100 ns, while PL and  Rice K-factor generally vary within narrower ranges of approximately 20 dB and 15 dB  respectively. 
Consequently, although prediction curves for PL and  Rice K-factor may exhibit significant deviations from ground truth, the associated increase in RMSE is relatively modest, ranging from 2-4 dB. 

In Fig. \ref{exp2}, despite  decreased accuracy of the models using  single segmented images compared to Fig. \ref{exp1}, the predicted trends for  PL, Rice K-factor, and RMS delay spread remain discernible. 
It indicates that the model trained on single segmented image datasets retains robust generalization performance and is capable of handling scenario migration effectively. 
In stark contrast, the models trained on   original and fully segmented images exhibit a significant degradation in prediction accuracy. 
These models fail to capture the trends of  three channel characteristics and even the fuzzy trends are not predicted. 
This highlights the poor generalization ability of the models trained with original and fully segmented images, which struggle to transfer to new scenarios.

\begin{figure}[!t]
	\centering
	\includegraphics[width=.5\textwidth]{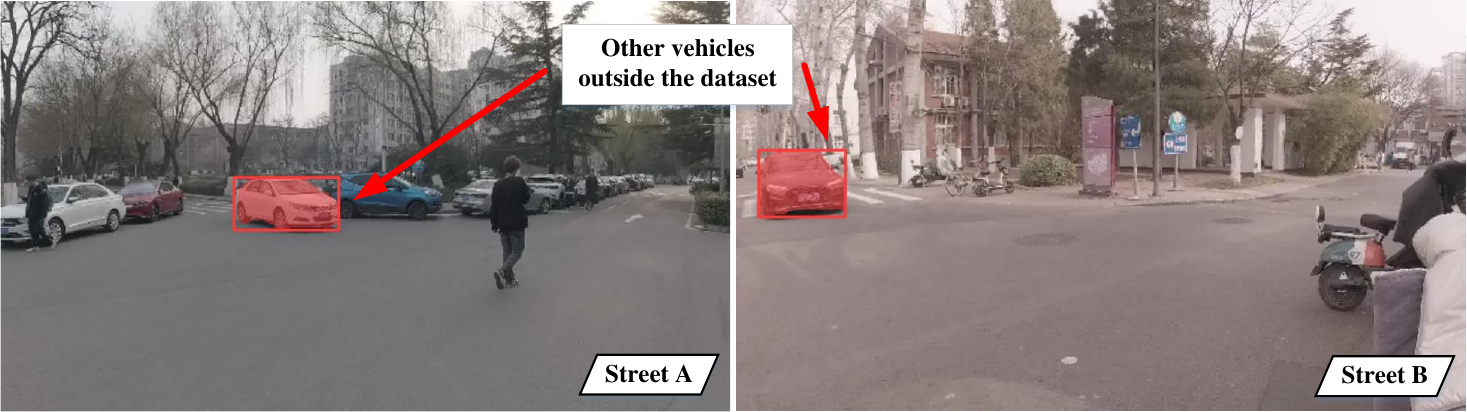}%
	\caption{Single segmented images of non-test vehicles.}                    
	\label{zache}
\end{figure}

\subsection{Experiment 3: Other Vehicle Cross-Validation}

\begin{figure*}[!t]
	\centering
	
	\subfloat[]{\includegraphics[width=.48\textwidth]{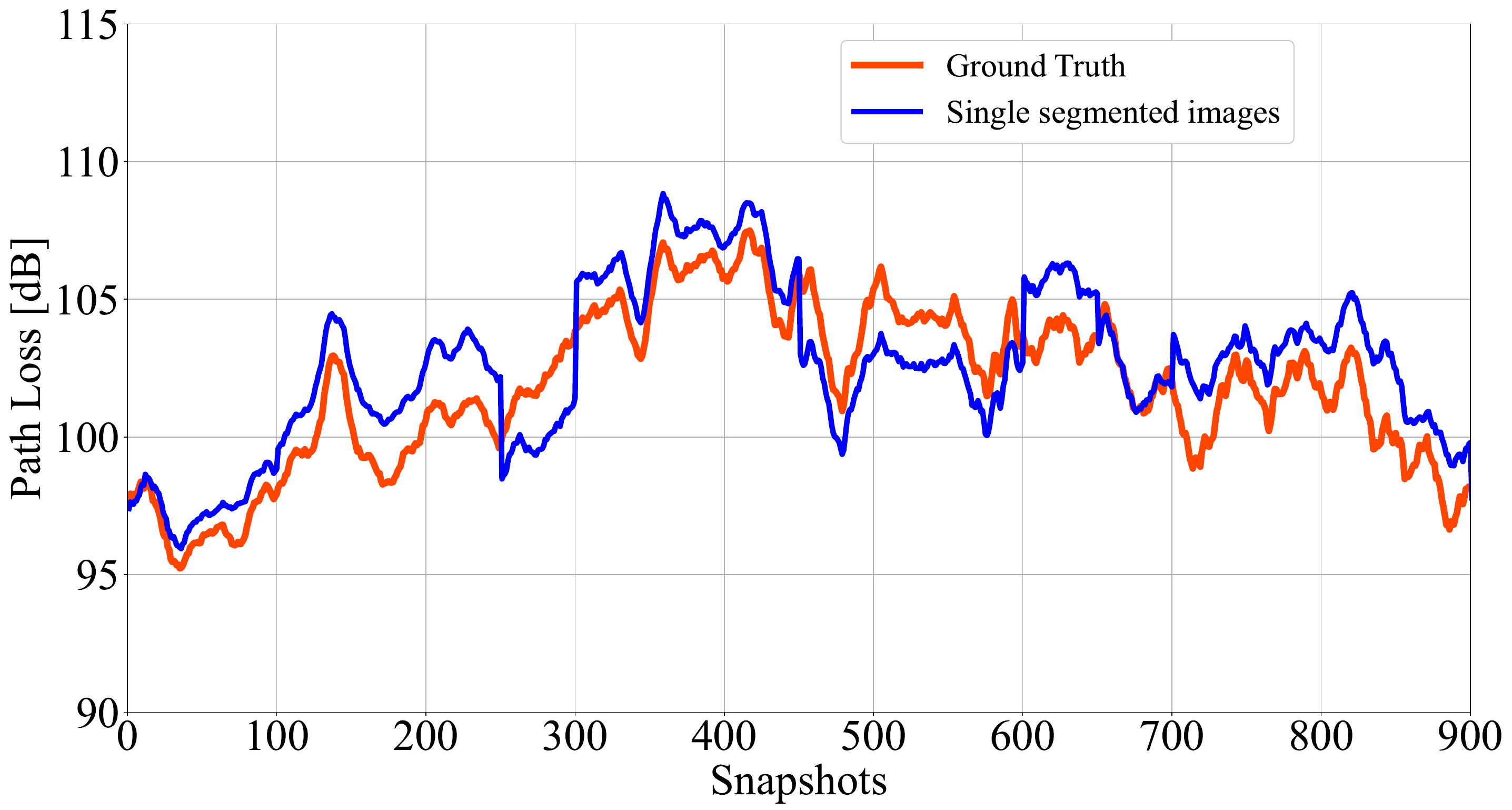}%
		\label{BaJiao_Day_1_PL_Exp3}}
	\subfloat[]{\includegraphics[width=.48\textwidth]{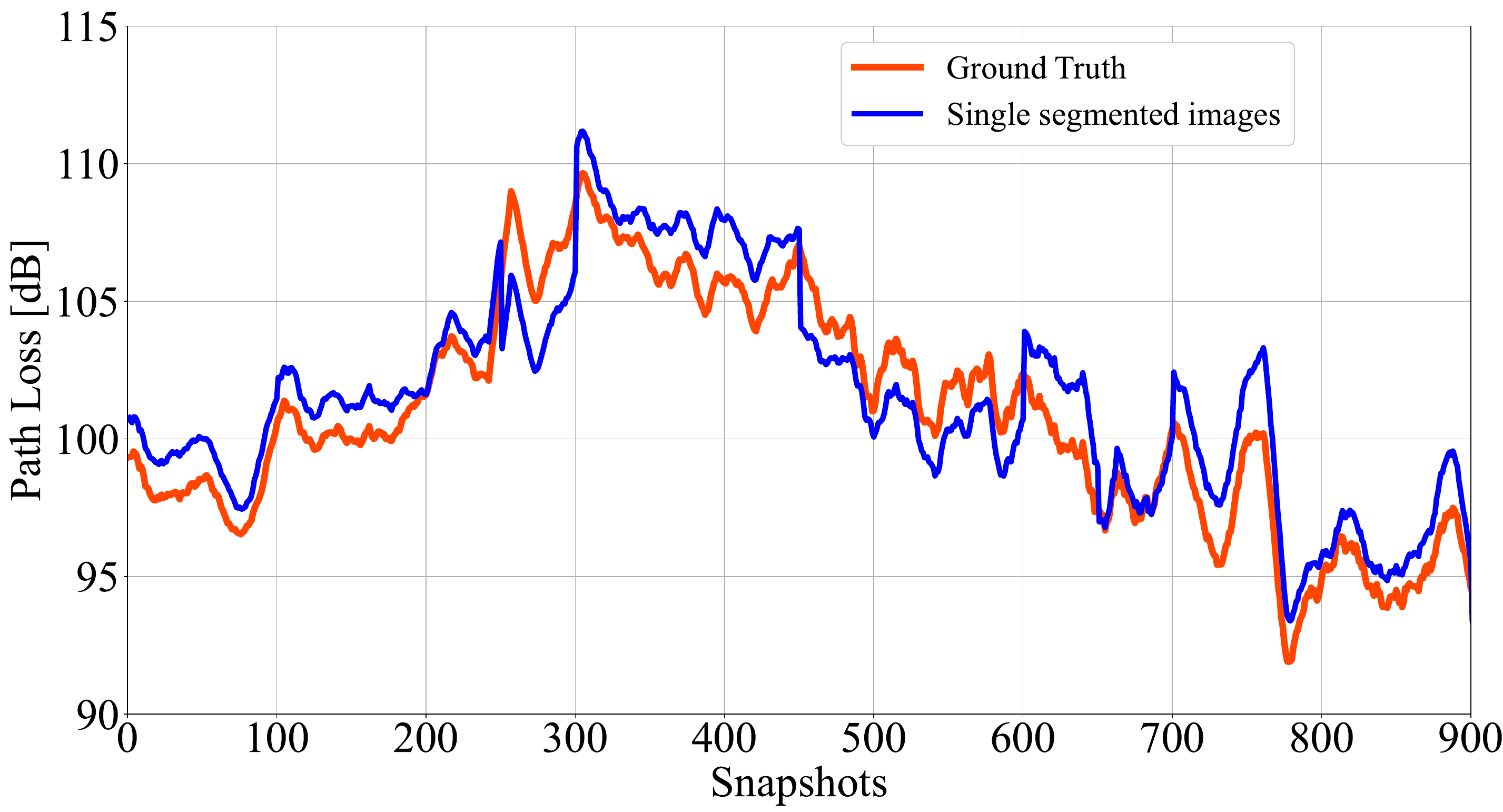}%
		\label{BaJiao_Day_1_DS_Exp33}}
	\quad
	\caption{
		Prediction results of PL in Experiment 3 using  single segmented images of other vehicles.
		(a) Street A;
		(b) Street B.
	}
	\label{exp3}
\end{figure*}

In this subsection, we introduce two  other vehicles to create new visual image test sets for further validation of the model's generalization capability.
Specifically, additional measurement campaigns are conducted on Streets A and B, where   Rx cabinet is mounted on two vehicles distinct from   Rx vehicle  depicted in Figs. \ref{measurement} and \ref{images}. 
As illustrated in Fig. \ref{zache}, the process for generating  image-channel datasets is exactly the same as mentioned in Section II-C. 
Subsequently, 900  continuous sequence of  snapshots are randomly extracted from   datasets of   two streets, with the target user in the visual data replaced by  these two other  vehicles.
The single segmented images   from Streets A and B are then input into Models A and B to predict PL, Rice K-factor, and RMS delay spread. 
The resulting prediction errors are presented in Table \ref{RMSE}.
Note that this experiment excludes the consideration of  original and fully segmented images. 
Taking PL as examples,  prediction results are shown in Fig. \ref{exp3}.

It can be found that when target user in RGB  images is replaced with vehicles of different shapes, colors and sizes, our models still demonstrate  strong prediction performance as shown in Fig. \ref{exp3}. 
Prediction errors for Models A and B in terms of PL, Rice K-factor, and RMS delay spread are 2.87–2.94 dB, 2.98–3.27 dB, and 3.18–3.45 ns, respectively. 
These RMSEs  are fairly close to those obtained in Experiment 1, where single segmented images are used, with the maximum difference not exceeding 2 dB.
It can be speculated that models A and B have effectively learned   channel characteristic parameters associated with  target user  represented by   mask and  rectangular bounding box at various positions in  RGB images (or streets), rather than relying on the number of pixel blocks within  masks.
The observed results indicate that when only  target user is segmented, prediction network exhibits robust generalization capabilities. 
The prediction accuracy shows minimal dependence on the shape, size, and  color  of  target user.

\subsection{Discussion}

To thoroughly evaluate the performance of proposed model, we have conducted three experiments to assess its capabilities.
First, we perform  scenario self-validation and cross-validation experiments on PL, Rice K-factor, and RMS delay spread using three types of images: single segmented images, original images, and fully-segmented images. 
The prediction results indicate that  models trained on the same street dataset with all three types of  image  are able to predict the trend of channel characteristics in  V2I scenario with minimal occlusion and predominantly LoS propagation. 
When using single segmented images, even subtle fluctuations can be  captured, while the other two types of images can only predict a rough fuzzy trend.
When  datasets of two streets are used for cross-validation, the model trained on single segmented images can still predict the changing trends, albeit with reduced accuracy. 
However, models trained on  original and fully segmented images failed to make accurate predictions altogether, with the RMSE increasing significantly. 
These models demonstrated a lack of generalization and an inability to adapt to different scenarios.

Finally, we replace  Rx vehicle in original single segmented image datasets with two vehicles of different shapes, sizes  and colors to form new test sets to test Models A and B respectively. 
Despite the changes in Rx vehicle appearance, we achieve  excellent prediction accuracy and acceptable RMSE, confirming that our models do  not rely on the specific appearance of  target user. 
The above results demonstrate that the proposed model exhibits high prediction accuracy across different streets and target users, showcasing remarkable generalization performance.

\section{CONCLUSION}
In this paper, we leverage the visual sensing data available at  BS, which does not consume additional spectrum resources, to propose a vision-aided channel prediction model suitable for   V2I scenarios.
The model requires only  RGB images of   target user captured at   BS, followed by instance segmentation of  target user within  images, and then inputting them   into the deep neural network, accurate prediction of channel characteristics can be achieved.
Specifically, we firstly  conduct  extensive measurement campaigns in   typical V2I scenarios, collecting channel and visual data from various streets. 
Image data consists of individual segmentation of  target user  using  YOLOv8 network. 
Channel data includes extracted channel characteristics including PL, Rice K-factor, and RMS delay spread.
Subsequently,  established dataset is used to train and test  prediction network, the ResNet-32. 
We perform  scenario self-validation and cross-validation experiments, using original and fully segmented images as comparisons to evaluate  model's performance.
The results indicate that models trained with single segmented images exhibit  excellent prediction accuracy across test sets from different streets, 
while models trained with the other images  lacks the ability to transfer to different streets.
Finally, we construct  new test sets by replacing target vehicle with other vehicles of   different appearances, and the test results still show  excellent prediction accuracy.
These findings demonstrate that  proposed model achieves high prediction accuracy and remarkable generalization performance across different streets and target users.
The ideas in this paper  will be  helpful to promote   further deep integration of AI (especially CV) technology and wireless communication and also provide a set of novel solutions for     realization of intelligent  vehicular communication system.
   
   \balance
   \bibliographystyle{IEEEtran}

   \nocite{*}
   
   \bibliography{IEEEabrv,ref}
\end{document}